\documentclass[letterpaper]{article} 
\usepackage{aaai23}  
\usepackage{times}  
\usepackage{helvet}  
\usepackage{courier}  
\usepackage[hyphens]{url}  
\usepackage{graphicx} 
\def\UrlFont{\rm}  
\usepackage{natbib}  
\usepackage{caption} 
\usepackage{algorithm}

\usepackage{newfloat}
\usepackage{listings}
\DeclareCaptionStyle{ruled}{labelfont=normalfont,labelsep=colon,strut=off} 
\floatstyle{ruled}
\newfloat{listing}{tb}{lst}{}
\floatname{listing}{Listing}
\newcommand{\BibTeX}{\rm B\kern-.05em{\sc i\kern-.025em b}\kern-.08em\TeX}
\usepackage{psfrag,amsmath,amsfonts,verbatim}
\usepackage{amsmath,amsfonts}
\usepackage{latexsym, amscd, amsfonts, eucal, mathrsfs, amsmath,amsthm, xypic,xr, stmaryrd, color, enumerate, mathtools}
\usepackage{amsmath,amsfonts}
\usepackage[utf8]{inputenc}

\usepackage{makecell}

\newcommand{\multilinecomment}[1]{}
\newcommand{\expectation}[4]{\mathbb{E}_{#1 \sim #2(\cdot | #3)}#4}

\DeclareSymbolFont{bbold}{U}{bbold}{m}{n}
\DeclareSymbolFontAlphabet{\mathbbold}{bbold}

\DeclareMathOperator*{\argmax}{arg\,max}

\newcommand{\corrpolicy}{\boldsymbol{\mu}}
\newcommand{\gencorr}{\boldsymbol{\mu}}
\newcommand{\gencorri}{\mu_i}
\newcommand{\gencorrj}{\mu_j}
\newcommand{\gencorrk}{\mu_k}

\newcommand{\act}{a}
\newcommand{\jointstrategy}{\boldsymbol{a}}
\newcommand{\fulltype}{\mathbf{t}}
\newcommand{\subtype}{\mathbf{s}}
\newcommand{\randomization}{\mathcal{C}}
\newcommand{\genpunish}{\boldsymbol{\hat{\tau}}}
\newcommand{\punishment}{\boldsymbol{\tau}}
\newcommand{\genpunishi}{\hat{\tau}_i}
\newcommand{\genpunishk}{\hat{\tau}_k}
\newcommand{\wrongguess}{c_{A,1}}
\newcommand{\exploited}{c_{A,2}}

\newcommand{\minusi}{_{-i}}
\newcommand{\minusj}{_{-j}}
\newcommand{\programspace}{P}
\newcommand{\prog}{p}
\newcommand{\fullprog}{\mathbf{p}}
\newcommand{\weakpoint}{v}
\newcommand{\attackguess}{\hat{v}}
\newcommand{\revelation}{\mathcal{R}}
\newcommand{\partialrevelation}{\mathcal{T}}
\newcommand{\attackprofit}{z}
\newcommand{\randdomain}{C}
\newcommand{\rand}{{c}}
\newcommand{\fulldevice}{\mathbf{d}}
\newcommand{\devspace}{D}

\newcommand{\progr}{p^r}
\newcommand{\progy}{p^y}
\newcommand{\progir}{p_i^r}
\newcommand{\progiy}{p_i^y}

\newcommand{\progky}{p_k^y}
\newcommand{\progrrec}[1]{p^{r,#1}}
\newcommand{\progyrec}[1]{p^{y,#1}}
\newcommand{\progirrec}[1]{p_i^{r,#1}}
\newcommand{\progjrrec}[1]{p_j^{r,#1}}
\newcommand{\progiyrec}[1]{p_i^{y,#1}}
\newcommand{\progjyrec}[1]{p_j^{y,#1}}

\newcount\Comments  
\definecolor{darkgreen}{rgb}{0,0.5,0}
\newcommand{\kibitz}[2]{\ifnum\Comments=1\textcolor{#1}{#2}\fi}

\usepackage{algorithm}
\usepackage[noend]{algpseudocode}

\usepackage{tikz}

\usetikzlibrary{bayesnet, arrows.meta,snakes}
\usetikzlibrary{matrix, fit}
\usetikzlibrary{decorations.markings}
\tikzset{
    myarrow/.style={
        -Stealth,
        shorten >=8pt,
        shorten <=8pt,
    },
    mymatrix/.style={
        matrix of math nodes
    },
}

\newtheorem{theorem}{Theorem}
\newtheorem{definition}{Definition}
\newtheorem{lemma}[theorem]{Lemma}
\newtheorem{proposition}{Proposition}

\theoremstyle{definition}
\newtheorem{exmp}{Example}

\newenvironment{sketch}{\paragraph{\normalfont \textsc{Proof Sketch.}}}{\hfill$\square$ \\}

\usepackage{mathtools}

\title{Commitment Games with
Conditional Information Disclosure}
\author{
    Anthony DiGiovanni\equalcontrib,
    Jesse Clifton\equalcontrib
}
\affiliations{
    Center on Long-Term Risk\\

    London, UK\\
    \{anthony.digiovanni, jesse.clifton\}@longtermrisk.org
}

\begin{document}

\maketitle

\begin{abstract}
The conditional commitment abilities
of
mutually transparent
computer agents
have been studied in previous work on 
commitment games
and 
program equilibrium.
This literature has shown how these abilities
can help resolve 
Prisoner’s Dilemmas and other failures of cooperation 
in complete information settings.
But
inefficiencies due to private information
have been neglected thus far in this 
literature, 
despite
the fact
that these 
problems are pervasive and
might also be addressed by greater mutual transparency. 
In this work, we introduce a framework for commitment games with
a new kind of conditional commitment device, which
agents can use to conditionally disclose private information. 
We prove a folk theorem for this setting  
that
provides sufficient conditions for ex post efficiency, 
and thus represents a model of ideal cooperation between agents 
without a third-party mediator.
Further, extending previous work on program
equilibrium, we develop
an implementation of conditional information
disclosure.
We show that this implementation forms program
$\epsilon$-Bayesian Nash equilibria
corresponding to the 
Bayesian Nash equilibria of these commitment games.
\end{abstract}

\section{Introduction}

What are the upper limits on the ability of 
rational, self-interested agents to cooperate?
As autonomous systems become increasingly responsible for  
important decisions,
including in interactions with other agents,
the study of ``Cooperative AI'' \citep{dafoe2020open} will potentially help
ensure these decisions result in cooperation.
It is well-known that game-theoretically rational behavior
--- which will potentially be more descriptive of
the decision-making of advanced computer agents than humans ---
can
result in imperfect cooperation, in the 
sense of inefficient outcomes. Some famous
examples are 
the Prisoner's Dilemma 
and the Myerson-Satterthwaite impossibility of
efficient bargaining under incomplete information
\citep{myerson1983efficient}. 
\citet{fearon1995rationalist} explores 
``rationalist'' explanations for war 
(i.e., situations in which war occurs in 
equilibrium); 
these include Prisoner's Dilemma-style 
inability to credibly commit to peaceful 
alternatives to war, as well as 
incentives to misrepresent
private information (e.g.,
military 
strength).
Because private information
is so ubiquitous in real strategic interactions,
resolving these cases of inefficiency
is a fundamental open problem.
Inefficiencies due to private information will be increasingly observed in machine learning, as machine learning is used to train agents in complex multi-agent environments featuring private information, such as negotiation.
For example, \citet{dealornodeal} found that when an agent was trained
with reinforcement
learning on negotiations under incomplete information,
it failed to reach an agreement with humans more frequently than a
human-imitative model did.

\par But
greater ability to make commitments
and share private information can 
open up more efficient equilibria. 
Computer
systems could be much better
than humans at making
their internal workings legible to other
agents, and at making sophisticated 
conditional commitments.
More mutually beneficial outcomes
could also be
facilitated by new technologies like
smart contracts \citep{varian2010computer}. 
This makes the game theory of 
interactions between agents with these abilities 
important for the understanding of
Cooperative AI
--- in particular, for developing an ideal
standard of multi-agent decision making
with future technologies. 
An extreme example of the power of 
greater 
transparency and commitment ability is 
\citet{tennenholtz2004program}'s ``program 
equilibrium'' solution to the one-shot
Prisoner's Dilemma. The players in 
Tennenholtz's ``program game'' version 
of the 
Prisoner's Dilemma
submit computer programs to play on their behalf,
which
can condition their outputs on 
each other's source code. 
Then a pair of programs with the 
source code \texttt{``If counterpart's source code
== my source code: Cooperate; Else: Defect''} 
form an equilibrium 
of mutual cooperation.

\par In this spirit, we are interested in exploring 
the kinds of cooperation that can be
achieved by agents who are capable 
of extreme mutual transparency and credible 
commitment.
We can think of this as giving 
an upper bound on the ability of
advanced artificially intelligent agents, or 
humans equipped with advanced technology
for commitment and transparency, to 
achieve efficient outcomes.
While such abilities are inaccessible
to current systems,
identifying sufficient conditions
for cooperation under private information
provides directions for future research
and development,
in order to avoid failures of cooperation.
These are our main contributions: 
\begin{enumerate}
    \item We develop a new class of games in which
    players can condition both their commitments and 
    disclosure of private information on their 
    counterparts' commitments and decisions to disclose 
    private information.
    We present a folk theorem for these games: 
    The set of equilibrium payoffs
    equals the set of feasible and interim 
    individually rational payoffs,
    notably including all ex post efficient payoffs. The 
    equilibria are conceptually straightforward: 
    For a given ex post payoff profile, 
    players disclose their private information
    and play according to an action 
    profile 
    attaining
    that payoff profile;
    if anyone deviates, 
    they revert to a punishment
    policy
    (without disclosing private
    information to
    the deviator).
    The problem is to avoid
    circularity in these conditional decisions.
    Our result
    builds on \citet{FORGES201364}' folk theorem 
    for Bayesian games without conditional information
    disclosure, in which equilibrium payoffs must also 
    be incentive compatible. 
    This expansion of the set of equilibrium payoffs is important,
    because in several settings,
    such as those of the classic Myerson-Satterthwaite theorem \citep{myerson1983efficient},
    ex post efficiency (or optimality according to some function of social welfare) and incentive compatibility are mutually exclusive.
    \item In these commitment games,
    the conditional commitment and 
    disclosure devices are abstract
    objects. 
    The devices in \citet{FORGES201364}' and our folk theorems
    avoid circularity by conditioning
    decisions 
    on the particular identities
    of the other players' devices,
    but this precludes robust cooperation
    with other devices that would output
    the same decisions.
    Using computer programs as conditional commitment
    and disclosure devices, 
    we give a specific implementation of 
    $\epsilon$-Bayesian Nash equilibria corresponding to the 
    equilibria
    of our commitment game. This approach
    extends 
    \citet{oesterheld2019robust}'s ``robust program equilibria.''
    We solve the additional problems of
    (1) ensuring that
    the programs terminate
    with more than two players,
    (2) in circumstances where
    cooperating with other players requires
    knowing their private information.
    Ours is the first study of program equilibrium \citep{tennenholtz2004program}
    under private information.
\end{enumerate}
%

\section{Related Work}\label{sec:related}

\paragraph{Commitment games and program equilibrium.}
We build on \textit{commitment games}, introduced 
by 
\citet{kalai2010commitment} and generalized to
Bayesian games (without verifiable
disclosure)
by \citet{FORGES201364}. In a commitment game, players submit commitment
devices that can choose actions conditional on other players' 
devices. 
This leads to folk theorems: 
Players can choose commitment devices that conditionally commit to 
playing a target action (e.g., cooperating in a Prisoner's Dilemma), 
and punishing if their counterparts do not play accordingly (e.g., 
defecting in a Prisoner's Dilemma if counterparts' devices do not cooperate).
A specific kind of commitment game is one played
between computer agents who can condition 
their behavior on each other's
source code. 
This is the focus of the literature on 
program equilibrium
\citep{rubinstein,tennenholtz2004program,lavictoire2014program,critch2019parametric,oesterheld2019robust,oesterheld2021safe}. 
\citet{Peters2012DefinableAC} critique the program equilibrium framework
as insufficiently robust to new contracts, 
because the programs in, e.g., \citet{kalai2010commitment}'s folk
theorem
only cooperate with the exact programs
used in the equilibrium profile.
Like ours, the commitment devices in \citet{Peters2012DefinableAC}
can disclose their types and punish
those that do not also disclose.
However, their devices disclose unconditionally
and thus leave the punishing player
exploitable, restricting the equilibrium payoffs to a smaller 
set than that of \citet{FORGES201364} or ours.

\par Our folk theorem  
builds directly on \citet{FORGES201364}. 
In Forges' setting, players
lack
the ability to disclose private 
information.
Thus the equilibrium payoffs
must be incentive compatible. 
We instead 
allow
(conditional) verification
of private information, which lets
us drop Forges' incentive compatibility 
constraint on equilibrium payoffs. 
Our program 
equilibrium implementation extends
\citet{oesterheld2019robust}'s robust program 
equilibrium to allow for conditional 
information disclosure.

\paragraph{Strategic information revelation.}
In games of strategic information revelation, 
players have the ability to verifiably 
disclose some or all of their private 
information. 
The question then becomes: How much
private information should players disclose (if any),
and how should
other players update their beliefs based on players'
refusal
to disclose some information? A foundational result in this
literature is that of \textit{full unraveling}: Under a range
of conditions, when players can verifiably disclose
information, they will act as if all information has been disclosed
\citep{milgrom1981good,grossman1981informational,milgrom1986relying}.
This means the mere possibility of verifiable disclosure
is often enough to avoid informational inefficiencies. 
However, there are cases where unraveling 
fails to hold and, even
when verifiable disclosure is possible,
informational inefficiencies persist and lead
to welfare losses.
This can be due to 
uncertainty about a player's ability to verifiably disclose
\citep{dye1985disclosure, shin1994burden}
or
disclosure being costly \citep{grossman1980disclosure,jovanovic1982truthful}.
But disclosure of private information can fail
even without
such uncertainty or costs
\citep{Kovenock09informationsharing,martini18}.
We will show how these kinds of private information
problems can be 
remedied
with the commitment technologies of our framework
(but not weaker ones, like
those of \citet{FORGES201364}).

\section{Preliminaries: Games of Incomplete Information and Inefficiency}
\label{sec:preliminaries}

\subsection{Definitions}\label{sec:definitions} 
Let $G$ be a Bayesian game with $n$ players. Each player~$i$ has a space of types $T_i$, giving a joint
type space $T = \bigtimes_{i=1}^n T_i$. At the start of the 
game, players' types are sampled by Nature according to the 
common
prior $q$. Each player knows only their type. 
Player $i$'s strategy is a choice of action $\act_i \in \mathcal{A}_i$
for each type in $T_i$.
Let $u_i(\fulltype, \jointstrategy)$ denote player $i$'s expected payoff in this game
when the players have types $\fulltype = (t_1, \dots, t_n)$ and
follow an action profile 
$\jointstrategy = (\act_1, \dots, \act_n)$. A Bayesian Nash equilibrium 
is an action profile $\jointstrategy$ in which every player 
and type plays a best response with respect to 
the 
prior over other players' types:
For all players $i$ and all types $t_i$, 
$\act_i(t_i) \in \argmax_{\act_i' \in \mathcal{A}_i} 
\expectation{\fulltype\minusi}{q}{t_i}{u_i(\fulltype, (\act_i'(t_i), \jointstrategy\minusi(\fulltype\minusi)) )}$.
An $\epsilon$-Bayesian Nash equilibrium is similar: Each 
player and type expects to gain at most $\epsilon$ 
(instead of 0) by deviating 
from $\jointstrategy$.

\par  
We assume players can correlate their actions
by conditioning on a trustworthy
randomization signal $\randomization$.\footnote{Even without a
trusted third party
that supplies a common correlation signal,
players could choose to
all condition on the same
“natural” source of randomness.}
For any correlated policy $\gencorr$ (a distribution 
over action profiles), let
$u_i(\fulltype,\gencorr) = \mathbb{E}_{\jointstrategy 
\sim \gencorr} u_i(\fulltype,\jointstrategy)$.
When it is helpful,
we will write $\gencorr(\cdot | \subtype)$
to clarify the subset of the type profile $\subtype \subseteq \fulltype$
on which the
correlated policy is conditioned.
Let
$(\act_j, \gencorr_{-j})$ denote a correlated policy
such that player $j$ plays $\act_j$
with probability 1,
and the actions of players other than $j$ are sampled from
$\gencorr_{-j}$ independently of $\act_j$.
Then, 
the following definitions will be key to 
our discussion:

\begin{definition} A payoff vector $\mathbf{x}$ as a function of type profiles is \textbf{feasible} if there is a correlated policy $\gencorr(\cdot|\fulltype)$
such that, for all players $j$ and types $t_j \in T_j$, $x_j(t_j) =
    \expectation{\fulltype_{-j}}{q}{t_j}{u_j(\fulltype, \gencorr)}$.
\end{definition}

\begin{definition}
A payoff $\mathbf{x}$ is \textbf{interim individually rational (INTIR)} if, for each player $j$, there is a correlated policy
$\punishment_{-j}(\cdot | \fulltype_{-j})$
used by the other players such that, for all $t_j \in T_j$, $x_j(t_j) \geq 
    \max_{\act_j \in \mathcal{A}_j} \expectation{\fulltype_{-j}}{q}{t_j}{u_j(\fulltype,(\act_j,
    \punishment_{-j}(\cdot | \fulltype_{-j})))}$.
\end{definition}

The
\textit{minimax policy}
$\punishment_{-j}$
is 
used by the other players to punish player $j$. 
The threat of such punishments will 
support the equilibria of our folk theorem. 
Players
only have sufficient information to 
use this correlated policy
if they disclose their types to each other. Moreover, 
the punishment can 
only work in general if they do \textit{not} disclose their 
types to player 
$j$, because
the definition of INTIR requires
the deviating player to be uncertain about
$\fulltype\minusj$.
Since the inequalities hold 
for all $t_j \in T_j$,
the players do not need to know player~$j$'s type to punish them. 

\begin{definition}
A feasible payoff $\mathbf{x}$ induced by $\gencorr$ is \textbf{incentive compatible (IC)} if, for each player $j$ and type pair $t_j, s_j \in T_j$, $x_j(t_j) \geq
    \expectation{\fulltype_{-j}}{q}{t_j}{u_j((t_j,\fulltype_{-j}), \gencorr(\cdot | s_j, \fulltype_{-j}))}$.
\end{definition}
\noindent Incentive compatibility means that,
supposing players report their
part of a type profile
on which their correlated policy
is conditioned,
no player prefers to lie about their type.


\begin{definition} Given a type profile $\fulltype$, a payoff $\mathbf{x}$ is \textbf{ex post efficient}
(hereafter, ``efficient'')
if there does not exist 
$\boldsymbol{\tilde \gencorr}$ such that $u_i(\fulltype, \boldsymbol{\tilde 
\gencorr}) \geq x_i(t_i)$ for all $i$ and $u_{i'}(\fulltype, \boldsymbol{\tilde 
\gencorr}) > x_{i'}(t_{i'})$ for some $i'$.
\end{definition}


We will also consider games with strategic
information revelation,
i.e., Bayesian games where,
immediately after
learning their types
and before playing $\jointstrategy$,
players 
can
disclose
their private information as follows (the ``disclosure phase'').
Players simultaneously each choose
$\Theta_i = (\Theta_{ij})_{j \neq i}$,
where each
disclosure set~$\Theta_{ij}$
is from some disclosure
space $\revelation(t_i)$,
a subset of
$\partialrevelation(t_i) = \{\Theta_i \subseteq T_i \ | \ t_i \in \Theta_i\}$.
Then, each player~$j$ observes each~$\Theta_{ij}$, thus learning that player $i$'s type is in $\Theta_{ij}$,
and conditions $\act_j$ on $(\Theta_{ij})_{i\neq j}$.
As is standard in the framework
of strategic information
revelation,
disclosure is \textit{verifiable} in the sense
that 
each $\Theta_{ij}$ must
contain player~$i$'s
true type;
they cannot falsely ``disclose'' a different type.
We will place our results on conditional type disclosure
in the context of the literature on unraveling:

\begin{definition}
Let $\{\Theta_i\}_{i=1}^n$ be the profile of disclosure 
set lists
(as functions of types) in
a Bayesian Nash equilibrium~$\sigma$ of a game with strategic
information revelation.
Then~$\sigma$
has \textbf{full unraveling} if
$\Theta_{ij}(t_i) = \{t_i\}$ for all $i,j$, or \textbf{partial unraveling} if
$\Theta_{ij}(t_i)$ is a strict subset of $T_i$ for
some
$i,j$.
\end{definition}

\subsection{Inefficiency: Motivating example}\label{sec:running-example}

Uncertainty about others' private information, and a
lack of ability or incentive to disclose that 
information, can lead to inefficient outcomes 
in Bayesian Nash equilibrium (or an 
appropriate refinement thereof). Here is an example 
we use to illustrate how informational 
problems can be overcome under our assumptions, 
but not under the weaker assumption of unconditional 
disclosure ability.

\begin{exmp}[War under incomplete information, adapted from \citet{slantchev2011mutual}]
\label{introexample}
Two countries $i=1,2$ are on the verge of war over some territory.
Country 1 offers a split of the territory
giving fractions $s$ and $1-s$ to countries
1 and 2, respectively.
If country 2 rejects this offer, they go to war.
Each player wins with some probability (detailed below), and each 
pays a cost of fighting $c_i > 0$. The winner
receives a payoff of 1, and the loser gets 0.

\par The countries' military strength determines the probability 
that country 2 wins the war, denoted $p(\theta)$.
Country~$1$ doesn't know 
whether country 2's army is weak (with type $\theta^W$)
or
strong ($\theta^S$), 
while country 1's strength is common knowledge.
Further, country 2 has a weak point, which country 1 believes
is equally likely to be in
one of two locations $\weakpoint \in \{1, 2\}$.
Thus country $2$'s type is given by 
$t_2 = \{\theta, \weakpoint\}$.
Country 1 can make a sneak attack on
$\attackguess \in \{1, 2\}$,
independent of whether they go to war.
Country 1 would gain $\attackprofit$ from
attacking $\attackguess = \weakpoint$,
costing $\exploited$ for country 2.
But incorrectly attacking $\attackguess \neq \weakpoint$ would
cost $\wrongguess > \attackprofit$ for country~1,
so country 1 would not risk an attack given
a prior of~$\frac{1}{2}$ on each of the locations.
If country 2 discloses its full type
by allowing inspectors from country 1 to assess
its military strength $\theta$,
country 1 will also learn $\weakpoint$. 

\par
If country 1
has a sufficiently low 
prior 
that country 2 is strong,
then war occurs in the unique
perfect Bayesian equilibrium when country 2
is strong. Moreover, this can happen 
even if the countries can 
fully disclose their private information to one
another. 
In other words, the unraveling
of private information does not occur, because player 2 will be made worse off 
if they allow player 1 to learn about their 
weak point
(see Appendix
\ref{app:sub:example}
for a formal argument).
Thus, unconditional disclosure
is not sufficient to allow
efficiency in equilibrium in 
this example,
motivating the use of the conditional
disclosure devices defined
in the next section.
\end{exmp}

Next, we formally introduce our framework for commitment
games with conditional information disclosure 
and present our folk theorem.

\section{Commitment Games with 
Conditional Information Disclosure}
\label{results}

\subsection{Setup}
\label{sec:sub:setup}
Players
are faced with a ``base game'' $G$, 
a Bayesian game with strategic information revelation
as defined in
Section~\ref{sec:definitions}.
In our framework, a commitment
game is a higher-level Bayesian
game in which the type distribution is the same as that of $G$,
and
strategies
are
devices that define mappings from other players'
devices
to actions and disclosure
in $G$ (conditional on one's type).
We assume $\{\{t_i\}, T_i\} \subseteq \revelation(t_i)$ for all
players $i$ and types $t_i$,
i.e., players are at least
able to disclose their exact types
or not disclose any new information.
They additionally have access to 
devices that can condition (i) their actions in 
$G$ and (ii) the 
disclosure of their private information
on other players' 
devices.
Upon learning their type $t_i$, player $i$
chooses a commitment device~$d_i$ from an
abstract space of devices $D_i$.
These devices 
are indices that,
based on $t_i$,
induce
a
\textit{response function} and a 
\textit{type disclosure function}
(as detailed below).
As in \citet{kalai2010commitment} and \citet{FORGES201364}, we will 
define these functions so as to allow players 
to condition their decisions on each other's 
decisions without circularity.

\par
Let $\randdomain$ be the domain
of the randomization signal $\randomization$ (a random variable),
and $\devspace_{-i} = \bigtimes_{j \neq i} \devspace_j$.
First, adopting the notation of \citet{FORGES201364},
the response function is 
$r^i_{d_i(t_i)}: \devspace_{-i} \times \randdomain
\to \mathcal{A}_i$.
Given the other players' devices $\fulldevice_{-i} = (d_j)_{j\neq i}$ and
the realized value~$\rand$ of
$\randomization$,
player~$i$'s action in~$G$
after the disclosure phase
is
$r^i_{d_i(t_i)}(\fulldevice_{-i},\rand)$.\footnote{A player who chooses to ``not commit'' submits 
a device that is not a function of the other players' devices. In 
this case, the other devices can only condition on this 
non-commitment choice, not on the particular action this player 
chooses.}
Conditioning the response on $\rand$
lets players commit to correlated distributions
over actions.

Second,
we introduce
type disclosure functions $y^i_{d_i(t_i)}: \devspace_{-i} \to 
\{0, 1\}^{n-1}$, which
are not in the framework
of \citet{FORGES201364}.
The $j$th entry of $y^i_{d_i(t_i)}(\fulldevice_{-i})$
indicates whether player $i$ discloses their type
to player~$j$, i.e.,
player~$j$ learns $\Theta_{ij} = \{t_i\}$ if this value is 1
or $\Theta_{ij} = T_i$ if it is~0.
(We can restrict
attention to cases where either all or no 
information is disclosed, as our folk
theorem shows that such a disclosure
space
is sufficient to enforce each equilibrium
payoff profile.)
Thus, each player~$i$ can
condition their action $r^i_{d_i(t_i)}(\fulldevice_{-i},\rand)$ on
the others' private information
disclosed to them via $(y^j_{d_j(t_j)}(\fulldevice_{-j}))_{j\neq i}$.
Further, they can choose whether to
disclose their type to another player, via $y^i_{d_i(t_i)}(\fulldevice_{-i})$,
based on that player's device.
Thus players can decide not
to disclose private information
to players whose devices
are not in a desired
device profile,
and instead punish them.

Then, the commitment game $G(\mathcal{D})$ is the one-shot Bayesian game
in which each player $i$'s strategy is a device $d_i \in D_i$, as a function of their type.
After
devices are simultaneously and independently submitted
(potentially as a draw from a mixed strategy over devices), 
the value~$\rand$ is drawn
from the randomization signal $\randomization$,
and
players play the induced action profile
$(r^i_{d_i(t_i)}(\fulldevice_{-i},\rand))_{i=1}^n$ in $G$.
Thus the ex post payoff of player $i$ in $G(\mathcal{D})$
from a device profile $\fulldevice = (d_i)_{i=1}^n$
is $u_i(\fulltype, (r^i_{d_i(t_i)}(\fulldevice_{-i},\rand))_{i=1}^n)$.

\subsection{Folk theorem}

Our folk theorem consists of two results:
First, any feasible and INTIR 
payoff can
be achieved in equilibrium (Theorem~\ref{eqconstruct}). 
As a special 
case of interest, then, any efficient payoff
can be attained in equilibrium.
Second, all
equilibrium payoffs in $G(\mathcal{D})$
are feasible and INTIR
(Proposition \ref{converse}).
The proof of Proposition \ref{converse}
is straightforward (see Appendix 
\ref{app:prop_proof}).

\begin{theorem}
\label{eqconstruct}
Let $G(\mathcal{D})$ be any commitment game.
For type profile $\fulltype$, let $\corrpolicy$ be a correlated
policy
inducing a feasible and INTIR payoff profile $(u_i(\fulltype,\corrpolicy))_{i=1}^n$.
Let $\genpunish$ be a punishment policy that is arbitrary
except, if $j$ is the only player with
$d'_j \neq d_j$,
let $\genpunish$ be the
minimax policy
$\punishment_{-j}$
against player~$j$.
Conditional on the signal $\rand$, let $\corrpolicy^{\rand}(\fulltype)$ be the
deterministic action profile,
called the target action profile,
given by
$\corrpolicy(\cdot|\fulltype)$,
and let $\genpunish^\rand$
be the deterministic action profile
given by $\genpunish$.
For all players $i$ and types $t_i$, let $d_i$ be such that:
\begin{align*}
    r^i_{d_i(t_i)}(\fulldevice_{-i}',\rand) &= \begin{cases}
    \gencorri^{\rand}(\fulltype),& \text{if } \fulldevice_{-i}' = \fulldevice_{-i}\\
    \genpunishi^{\rand},& \text{otherwise,} \\
\end{cases} \\
    {y^i_{d_i(t_i)}(\fulldevice_{-i}')}_j &= \begin{cases}
    1,& \text{if } d'_{j} = d_j\\
    0,& \text{otherwise.} \\
\end{cases}
\end{align*}
Then, the device profile $\fulldevice$ is a Bayesian Nash equilibrium of $G(\mathcal{D})$.
\end{theorem}

\begin{proof} 
We first need to check that the
response and type disclosure functions only
condition on information available to the players.
If all players use $\fulldevice$, then by
construction of $y^i_{d_i(t_i)}$ they all disclose their types to
each other, and so are able to play 
$\corrpolicy(\cdot|\fulltype)$ 
conditioned on their type profile
(regardless of whether the induced payoff is IC).
If at least one player uses some other device,
the players who do use $\fulldevice$ still share their types with
each other, thus they can play $\genpunish$.

Suppose player $j$ deviates from $\fulldevice$. That is, player~$j$'s strategy in $G(\mathcal{D})$ is
$d_j' \neq d_j$.
Note that the outputs of player~$j$'s response
and type disclosure functions induced by
$d_j'$ may in general be the same as those
returned by $d_j$.
We will show that~$\genpunish^\rand$
punishes deviations from
the
target action profile regardless of these outputs,
as long as there is a deviation in \textit{functions} $r'^j$ or $y'^j$.
Let $\act'_j = r^j_{d_j'(t_j)}(\fulldevice_{-j},\rand)$. Then:
\begin{align*}
    &\mathbb{E}_{\fulltype_{-j} \sim q(\cdot | t_j)}(u_j(\fulltype, (\act'_j,\genpunish))|d'_j,\fulldevice_{-j}) \\
    &\qquad = \expectation{\fulltype_{-j}}{q}{t_j}{u_j(\fulltype,(\act_j',
   \punishment_{-j}(\cdot | \fulltype_{-j})))}\\
    &\qquad \leq \mathbb{E}_{\fulltype_{-j} \sim q(\cdot | t_j)}u_j(\fulltype,\corrpolicy) \tag{by INTIR} \\
    &\qquad = \mathbb{E}_{\fulltype_{-j} \sim q(\cdot | t_j)}(u_j(\fulltype, \corrpolicy)|d_j,\fulldevice_{-j}).
\end{align*}
This last expression is the
ex interim payoff of the proposed commitment $d_j$ given that the other players use $\fulldevice_{-j}$, therefore $\fulldevice$ is a Bayesian Nash Equilibrium.
\end{proof}

\begin{proposition}
\label{converse}
Let $G(\mathcal{D})$ be any commitment game.
If a device profile $\fulldevice$ is a Bayesian Nash equilibrium of $G(\mathcal{D})$, then the induced payoff $\mathbf{x}$ is feasible and INTIR.
\end{proposition}

Our assumptions
do not imply the equilibrium
payoffs are IC (unlike \citet{FORGES201364}).
Suppose a player $i$'s payoff would increase if the players
conditioned the correlated policy
on a different type (i.e., not IC). This does not
imply that a profit is possible by deviating from the equilibrium,
because in our setting the other players' actions are
conditioned on the type disclosed by $i$. In particular, as in our proposed
device profile, they may choose to play their part of the target action profile
only if all other players' devices
disclose their (true) types.

The assumptions that give rise to this class of commitment games with conditional information disclosure
are stronger than the ability to 
unconditionally disclose private information. 
Recalling the unraveling results from
Section \ref{sec:related}, unconditional disclosure
ability is sometimes sufficient  
for the full disclosure of private information,
or for disclosure of the information that prohibits incentive compatibility,
and thus the possibility of 
efficiency in 
equilibrium. But this is not always true, whereas
efficiency is always attainable in equilibrium
under our assumptions. 
In Appendix 
\ref{app:efficiency},
we
first show that full unraveling
fails in our motivating example when
country 2 has
a weak point.
Then, we discuss conditions under which the ability
to partially disclose private information
is sufficient for efficiency, and examples
where these conditions don't hold. 

\section{Implementation of Conditional Type Disclosure via Robust Program Equilibrium}
\label{sec:robust}

Having shown that
all efficient payoff profiles
are achievable in equilibrium
using
conditional commitment
and disclosure devices,
we next consider how players can
practically (and more robustly)
implement these abstract devices.
In particular, can players achieve
efficient equilibria without using
the exact device profile in Theorem~\ref{eqconstruct},
which can only cooperate with itself?
We now develop an implementation showing
that this is possible,
after providing some background.

\citet{oesterheld2019robust} considers two computer
programs playing a game. Each program can simulate 
the other in order to choose an action in
the game. He constructs a program equilibrium --- 
a pair of programs that form an equilibrium
of this game --- using ``instantaneous 
tit-for-tat'' strategies. In the Prisoner's Dilemma, 
the pseudocode
for these programs (called ``$\epsilon$\texttt{GroundedFairBot}'')
is: ``\texttt{With small probability~$\epsilon$: 
Cooperate; Else: do what my 
counterpart does when playing against me}.'' 
These programs
cooperate with each other and
punish
defection.
Note that these programs are recursive, but guaranteed
to terminate because of the $\epsilon$ probability that
a program outputs Cooperate unconditionally.  

\par We use this idea to implement conditional 
commitment and disclosure devices. For us, ``disclosing
private information and playing according to the target
action profile''
is analogous
to cooperation  
in the construction of 
$\epsilon$\texttt{GroundedFairBot}. Thus,
instead of a particular device profile
in which a device
cooperates
if and only if all other devices are in that profile,
we consider programs
that cooperate if and only
if all other programs output cooperation
against each other.
We will first describe the appropriate class of programs
for \textit{program games} under private information.
Then we develop our program, \texttt{$\epsilon$GroundedFairSIRBot}
(where ``SIR'' stands for
``strategic information revelation''),
and show that it forms a $\delta$-Bayesian Nash equilibrium
of a program game. Pseudocode for
\texttt{$\epsilon$GroundedFairSIRBot} is given in Algorithm \ref{alg:robustprogctr_full}. 

As in Section \ref{sec:sub:setup}, 
there is a base game $G$, and players
choose strategies that implement actions in $G$
conditional on each other's strategies.
In a program game, programs fill the
role of devices in a commitment game.
Player $i$'s strategy in the program game is
a choice $p_i$ from the program space $\programspace_i$, 
a set
of 
computable
functions from
$\bigtimes_{j=1}^n \programspace_j \times \randdomain \times \{0, 1\}$
to $\mathcal{A}_i \cup \{0, 1\}^{n-1}$.
A program returns either an action
or a type disclosure vector
(just as a device induces response
and type disclosure functions,
which return actions and disclosure vectors,
respectively).
Each program takes as input the players' program
profile,
the signal~$\rand$,
and a boolean
that equals~1 if the program's
output is an action, and~0 otherwise.
For brevity, we write $\progir$
for a call to a program with the boolean set to~$1$,
otherwise $\progiy$.
Letting $\fullprog = (\prog_m)_{m=1}^n$ be the players' program profile,
player~$i$'s
action
in $G$ is
a call to their program $p_i(\fullprog, \rand, 1)$.
(We refer to these initial program calls as the \textit{base calls}
to distinguish them from calls made by other
programs.)
Then, 
the ex post payoff of player $i$ in the program
game is $u_i(\fulltype, (p_j(\fullprog,
\rand, 1))_{j=1}^n)$.

Like \citet{oesterheld2019robust}, we will use
programs that
unconditionally terminate with some small probability.
To generalize this idea to a setting
with private information and more than
two players,
we now introduce some additional elements of the
program game and our program.
First, in addition to 
$\randomization$ in the base game,
there is a randomization signal
$\randomization^\programspace$ on which
programs can
condition their outputs.
By using $\randomization^\programspace$ to correlate decisions
to unconditionally terminate,
our program profile will be able to
terminate with probability~1,
despite the exponentially increasing number of
recursive program calls.
In particular,
$\randomization^\programspace$ reads
the call stack of the players' program profile.
At each depth level~$L$ of recursion reached in the
call stack,
a variable $U_L$ is independently
sampled
from $\text{Unif}[0,1]$.
Each program call at level~$L$
can read off the values of $U_L$ and $U_{L+1}$
from~$\randomization^\programspace$.
The index $L$ itself is not revealed, however, because programs that ``know'' they are
being simulated could
defect in the base calls, while
cooperating in simulations to deceive
the other programs.
Second, let \texttt{$\epsilon$GroundedFairSIRBot}$^r$ and \texttt{$\epsilon$GroundedFairSIRBot}$^y$
be calls to the program \texttt{$\epsilon$GroundedFairSIRBot}
with $\texttt{output\_action}= 1$ and  $\texttt{output\_action} = 0$, respectively.
To ensure that our programs terminate in play
with a deviating program,
\texttt{$\epsilon$GroundedFairSIRBot}$^r$
will call truncated
versions of its counterparts' disclosure programs:
For $\prog_i \in P_i$, let $[\prog_i]$ denote~$\prog_i$ with immediate termination upon
calling another program.

\begin{algorithm}[ht]
\caption{\texttt{$\epsilon$GroundedFairSIRBot}}
\label{alg:robustprogctr_full}
\begin{algorithmic}[1]
\Require Program profile $\fullprog$, randomization signal value $\rand$, boolean \texttt{output\_action} 
\If{\texttt{output\_action} $= 1$}
    \If{$U_{L+1} \geq \epsilon$} \label{line:high_prob}
        \For{$k \neq i$} \Comment{Check if each player discloses} \label{line:check_revelation}
        \State $\mathbf{y}^k \leftarrow p_k(\fullprog, \rand, 0)$
    \EndFor
    \If{$\mathbf{y}^k = \mathbf{1}$ for all $k \neq i$} \label{line:fulltypecheck}
        \If{$U_L < \epsilon$} \Comment{Unconditionally cooperate \!\!} \label{line:randomcoop}
            \State \Return $\gencorri^{\rand}(\fulltype)$ \label{line:act_randomcoop}
        \EndIf
        \For{$k \neq i$} \Comment{Check if others cooperate} \label{line:loop}
            \State $\act_k \leftarrow p_k(\fullprog, \rand, 1)$
        \EndFor
        \If{$\act_k = \gencorrk^{\rand}(\fulltype)$ for all $k \neq i$}
            \State \Return $\gencorri^{\rand}(\fulltype)$ \label{line:conditional_coop}
        \EndIf
    \EndIf
    \State \Return $\genpunishi^{\rand}$ \Comment{Punish given known $(\mathbf{y}^m)_{m \neq i}$} \label{line:punish}
    \EndIf
    \For{$k \neq i$} \Comment{Check truncated $\progky$} \label{line:first_checktruncated}
            \State $\mathbf{y}^k \leftarrow [p_k](\fullprog, \rand, 0)$ \label{line:checktruncated}
        \EndFor
        \If{$\mathbf{y}^k = \mathbf{1}$ for all $k \neq i$} \Comment{Full type known} \label{line:trunc_type_check}
            \State \Return $\gencorri^{\rand}(\fulltype)$ \label{line:trunc_type}
        \EndIf
        \State \Return $\genpunishi^{\rand}$ \label{line:trunc_punish}
\Else
    \State $\mathbf{y}^i \leftarrow \mathbf{0}$
    \If{$U_L < \epsilon$} \Comment{Unconditionally disclose} \label{line:rev_uncon}
        \State \Return $\mathbf{1}$  \label{line:unconrev}
    \EndIf
    \For{$k \neq i$} 
        \State $\mathbf{y}^k \leftarrow p_k(\fullprog, \rand, 0)$
        \State $\act_k \leftarrow p_k(\fullprog, \rand, 1)$
    \EndFor
    \If{$\mathbf{y}^k = \mathbf{1}$ and $\act_k = \gencorrk^{\rand}(\fulltype)$ for all $k \neq i$} \label{line:check_for_coop_1}
            \State \Return $\mathbf{1}$ 
    \EndIf
    \For{$k \neq i$}
        \If{$\mathbf{y}^k_i = 1$ and ($\act_k = \genpunishk^{\rand}$ or $U_{L+1} < \epsilon$)} \label{line:punish_rev}
            \State $\mathbf{y}^i_k \leftarrow 1$
        \EndIf
    \EndFor
    \State \Return $\mathbf{y}^i$
\EndIf
\end{algorithmic}
\end{algorithm}

\tikzstyle{every pin edge}=[<-,shorten <=1pt,snake=snake,line before snake=4pt]

\begin{figure}[ht]
\centering
\begin{tikzpicture}[pin distance=9.2mm]
    \node[latent, ellipse](r1) {$\progirrec{1}$};
    \node[obs, rectangle, below=of r1, xshift=-1cm, yshift=0.8cm](y) {$y^{j} = 1$?};
    \node[obs, rectangle, below=of y, xshift=-1cm, yshift=0.55cm](u) {$U_1 \geq \epsilon$ \texttt{and} $U_2 \geq \epsilon$?};
    \node[obs, rectangle, below=of u, xshift=-1cm, yshift=0.55cm](p) {$\act_{j} = \gencorrj^{\rand}(\fulltype)$?};
    \node[latent, ellipse, below=of p, xshift=2.75cm, yshift=0.85cm](r2) {$\progjrrec{2}$};
    \node[latent, ellipse, below=of p, xshift=4.25cm, yshift=0.85cm](y2) {$\progjyrec{2}$};
    \node[latent, ellipse, below=of r2, xshift=-2.75cm, yshift=0.7cm](r3a) {$\progirrec{3}$};
    \node[latent, ellipse, below=of r2, xshift=-1cm, yshift=0.7cm](y3a) {$\progiyrec{3}$};
    \node[latent, ellipse, below=of y2, xshift=-0.75cm, yshift=0.7cm](r3b) {$\progirrec{3}$};
    \node[latent, ellipse, below=of y2, xshift=1cm, yshift=0.7cm](y3b) {$\progiyrec{3}$};
    \node[obs, rectangle, below=of y3a, xshift=-1cm, yshift=0.7cm](u2) {$U_3 \geq \epsilon$?};
    \node[obs, rectangle, below=of u2, xshift=-1cm, yshift=0.55cm](p2) {\makecell[l]{$y^{j} = 1 \texttt{ and }$ \\ $(\act_{j} = \gencorrj^{\rand}(\fulltype)$ \\
    $\texttt{or } \act_{j} = \genpunish^{\rand}_{j}$ \\
    $\texttt{or } U_4 < \epsilon)$?}};
    \node[below=of p, xshift=-1cm, yshift=0.55cm](t1){$\gencorri^{\rand}(\fulltype)$};
    \node[below=of p, xshift=0.85cm, yshift=0.55cm](pun1){$\genpunishi^{\rand}$};
    \node[below=of u, xshift=1cm, yshift=0.55cm](t2){$\gencorri^{\rand}(\fulltype)$};
    \node[below=of y, xshift=1.1cm, yshift=0.55cm](pun2){$\genpunishi^{\rand}$};
    \node[below=of u2, xshift=1cm, yshift=0.55cm](pun3){$1$};
    \node[below=of p2, xshift=-1cm, yshift=0.55cm](out1){$1$};
    \node[below=of p2, xshift=1cm, yshift=0.55cm](out2){$0$};
    \node[below=of r3a,xshift=-1cm,yshift=0.41cm,pin=above right:{}](name){};
    \node[below=of r3b,xshift=-1cm,yshift=0.41cm,pin=above right:{}](name){};
    \node[below=of y3b,xshift=-1cm,yshift=0.41cm,pin=above right:{}](name){};
    \node[latent, ellipse, below=of p2, xshift=2cm, yshift=0.85cm](r4a) {$\progjrrec{4}$};
    \node[latent, ellipse, below=of p2, xshift=3.75cm, yshift=0.85cm](y4a) {$\progjyrec{4}$};
    \node[below=of r4a,xshift=-1cm,yshift=0.41cm,pin=above right:{}](name){};
    \node[below=of y4a,xshift=-1cm,yshift=0.41cm,pin=above right:{}](name){};
    \edge[color=lightgray] {r2} {r3a,y3a}
    \edge[color=lightgray] {y2} {r3b,y3b}
    \edge {r1} {y}
    \edge[color=lightgray] {p} {r2}
    \edge {y} {u,pun2}
        \node[below=of y, xshift=-1cm, yshift=0.98cm](label1){yes};
        \node[below=of y, xshift=1.05cm, yshift=0.98cm](label3){no};
    \edge[color=lightgray] {y} {y2}
    \edge {p} {t1,pun1}
        \node[below=of p, xshift=-1.1cm, yshift=0.9cm](label1){yes};
        \node[below=of p, xshift=0.15cm, yshift=0.9cm](label2){no};
    \edge {u} {p,t2}
        \node[below=of u, xshift=-1cm, yshift=0.98cm](label1){yes};
        \node[below=of u, xshift=1cm, yshift=0.98cm](label2){no};
    \edge {y3a} {u2}
    \edge {u2} {p2,pun3}
        \node[below=of u2, xshift=-0.875cm, yshift=0.95cm](label1){yes};
        \node[below=of u2, xshift=0.25cm, yshift=0.95cm](label2){no};
    \edge {p2} {out1,out2}
        \node[below=of p2, xshift=-1.2cm, yshift=0.95cm](label1){yes};
        \node[below=of p2, xshift=0.3cm, yshift=0.95cm](label2){no};
    \edge[color=lightgray]{p2} {r4a, y4a}
\end{tikzpicture}
\caption{Flowchart for a 2-player program game between player $i$ using \texttt{$\epsilon$GroundedFairSIRBot}, and player~$j$ using an arbitrary program. An edge to a white node indicates a call to the program in that node; to a gray node indicates a check of the condition in that node; and to a node without a border indicates the output of the most recent parent white node. Wavy edges
depict a call to the program in the parent node, with its child nodes omitted for space. Superscripts indicate the level of recursion.}
\label{fig:flowchart}
\end{figure}
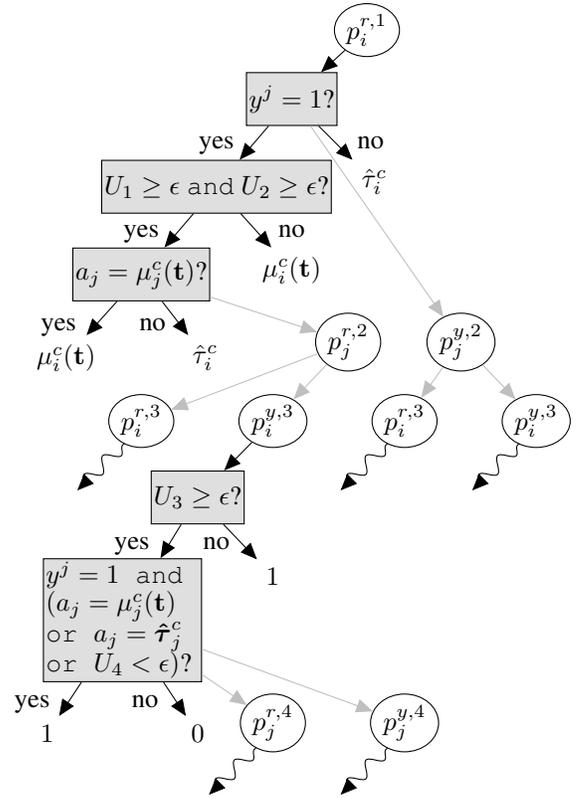

Figure \ref{fig:flowchart}
visually summarizes a program game
between \texttt{$\epsilon$GroundedFairSIRBot}
and some other program.
Like a device in the profile in Theorem \ref{eqconstruct},
which checks if the other devices are part of a profile
that disclose their types
and play their parts of the target action profile (``cooperate''),
our program checks if the other programs disclose and cooperate with it.
With high probability,
\texttt{$\epsilon$GroundedFairSIRBot}$^r$
checks if all other players' programs
disclose their types (lines \ref{line:high_prob}-\ref{line:fulltypecheck} of Algorithm~\ref{alg:robustprogctr_full}). If so, either with a small probability it
unconditionally cooperates (lines \ref{line:randomcoop}-\ref{line:act_randomcoop}),
or it cooperates only when all other programs cooperate
(lines~\ref{line:loop}-\ref{line:conditional_coop}).
Otherwise, it punishes (line~\ref{line:punish}).
If, with low probability,
the next call to \texttt{$\epsilon$GroundedFairSIRBot}$^r$
will unconditionally cooperate,
then the current call cooperates if and only if the other \textit{truncated} programs
disclose (lines \ref{line:first_checktruncated}-\ref{line:trunc_punish}).

In turn, \texttt{$\epsilon$GroundedFairSIRBot}$^y$
discloses its type unconditionally with probability $\epsilon$ (lines \ref{line:rev_uncon}-\ref{line:unconrev}).
Otherwise, it discloses to a given player $j$ under
two conditions (lines~\ref{line:check_for_coop_1} and \ref{line:punish_rev}).
First, player $j$ must disclose to the user.
Second, they must play an action consistent with
the desired equilibrium, i.e., cooperate when all
players disclose their types, 
or punish otherwise.

Unconditionally disclosing one's type and playing
the target action
avoids an infinite regress.
Crucially, these unconditional cooperation
outputs are correlated via~$\randomization^\programspace$.
Therefore, in a profile of copies of
this program,
either all copies unconditionally cooperate
together,
or none of them do so.
Using this property,
we can show (see proof of Theorem~\ref{robustprogeq} in Appendix
\ref{app:proof})
that a profile where all players use this program
outputs the target action profile with certainty. 
If one player deviates, first, \texttt{$\epsilon$GroundedFairSIRBot}$^r$ immediately
punishes if that player does not disclose.
If the deviating player does
disclose,
with some small probability the other players unconditionally cooperate (lines \ref{line:first_checktruncated}-\ref{line:trunc_type}),
making this strategy slightly exploitable,
but otherwise the deviator is punished.
Even if a deviation is punished,
\texttt{$\epsilon$GroundedFairSIRBot}$^y$ may unconditionally disclose.
In our approach, this margin of exploitability is the price of
implementing 
conditional commitment and disclosure 
with programs that cooperate based on counterparts' outputs,
rather than a strict matching of devices,
without an infinite loop.
Further, since a player is only able to
unconditionally cooperate under incomplete information if they know all players' types,
\texttt{$\epsilon$GroundedFairSIRBot}$^r$
needs to
prematurely terminate calls to programs that 
don't immediately unconditionally cooperate, but which 
may otherwise cause infinite recursion (line \ref{line:checktruncated}).
This comes at the expense of robustness: 
\texttt{$\epsilon$GroundedFairSIRBot}
punishes some players who may
have otherwise cooperated, with low probability.

\begin{theorem}
\label{robustprogeq}
Consider the program game induced by a base game $G$ and the program spaces $\{\programspace_i\}_{i=1}^n$.
Assume all strategies returned by these programs are computable.
For a type profile $\fulltype$, let $\corrpolicy(\cdot|\fulltype)$ induce a feasible and INTIR payoff profile $(u_i(\fulltype,\corrpolicy))_{i=1}^n$. 
Let~$\genpunish$
be 
the minimax policy
if one player~$j$ deviates,
and arbitrary otherwise.

Let $\overline{u}$ be the maximum payoff achievable by any player in~$G$, and
$\delta = \overline{u}((1-\epsilon)^{-2} - 1)$.
Then the program profile~$\fullprog$ given by Algorithm \ref{alg:robustprogctr_full} (with \texttt{output\_action} $= 1$) for players $i=1,\dots,n$ is a $\delta$-Bayesian Nash equilibrium. That is,
if players $i \neq j$ play this profile, and player $j$ plays a program $p_j' \in P_j$
that terminates with probability 1 given that any programs
it calls terminate with probability 1, then:
\begin{align*}
    & \mathbb{E}_{\fulltype_{-j} \sim q(\cdot | t_j)}(u_j(\fulltype, \corrpolicy)|\prog_j',\fullprog_{-j}) \\
    &\qquad \leq \delta + \mathbb{E}_{\fulltype_{-j} \sim q(\cdot | t_j)}(u_j(\fulltype, \corrpolicy)|\prog_j,\fullprog_{-j}).
\end{align*}
\end{theorem}

\begin{sketch}
We need to check (1) that the program profile $\fullprog$ terminates
(a) with or (b) without a deviation,
(2) that everyone plays the target action profile
when no one deviates,
and (3) that with high probability
a deviation is punished.
First, suppose no one deviates.
If $U_L < \epsilon$ for two levels of recursion in a row,
the calls to $\progiy$ and $\progir$
all unconditionally disclose 
(line \ref{line:unconrev})
and output the
target action (line \ref{line:trunc_type}),
respectively.
Because these unconditional cooperative outputs
are correlated through $\randomization^\programspace$,
the probability that $U_L < \epsilon$
at each pair of subsequent levels in the call stack
is a nonzero constant. Thus it is guaranteed to occur eventually and
cause termination in finite time, satisfying (1b).
Moreover, each call to $\progiy$
or $\progir$ in previous
levels of the stack sees that the next
level cooperates, and thus cooperates as well,
ensuring that the base calls
all output the target action profile. This shows (2).

If, however, one player deviates, we use the same guarantee
of a run of subsequent $U_L < \epsilon$ events
to guarantee termination. First, all calls to non-deviating programs terminate,
because any call to \texttt{$\epsilon$GroundedFairSIRBot}$^r$ conditional on $U_{L+1} < \epsilon$
forces termination (line \ref{line:checktruncated}) of calls to other
players' disclosure programs.
Thus the deviating programs also terminate, since they call terminating non-deviating programs.
This establishes (1a).
Finally, in the high-probability event that
the first two levels of calls to $\fullprog$ do
\textit{not} unconditionally cooperate,
\texttt{$\epsilon$GroundedFairSIRBot}$^r$
punishes the deviator as long as they do not
disclose their type and play their target action.
The punishing players will know each
other's types, since a call to \texttt{$\epsilon$GroundedFairSIRBot}$^y$ is
guaranteed by line~\ref{line:punish_rev} to disclose to anyone
who also punishes or \textit{unconditionally}
cooperates in the next level. Condition (3) follows.
\end{sketch}

We now discuss two practical considerations for this program
equilibrium implementation.
First, one obstacle to this implementation is 
demonstrating to one's counterpart that one's 
behavior is actually governed by 
the source code that has been shared.
In our program game with private information, 
there is the additional problem
that, as soon as one's source code is shared, 
one's counterpart may be able to 
read off one's private information 
(without disclosing their own). Addressing
this in practice might involve modular
architectures, where players could expose
the code governing their strategy without
exposing the code for their private
information. Alternatively, 
consider
AI
agents that can
place copies of themselves in a
secure box, where the copies can
inspect each other's full code but
cannot take any actions outside the
box.
These copies read each other's commitment devices off of their source code,
and report the
action and type outputs of these devices 
to the original agents.
If any copy within the box attempts to transmit information
that another agent's device refused to disclose, the box deletes its contents.
This protocol does not require a mediator or arbitrator; the agents and their copies make all the
relevant strategic decisions,
with the box only serving as a security mechanism.
Applications of secure multi-party computation
to machine learning \citep{knott2021crypten}, or
privacy-preserving smart contracts \citep{smartcontract} — with the
original agents treated as the ``public'' from whom
code shared among the copies is kept private
— might
facilitate the implementation
of our proposed commitment devices.

Second, it is an open question how to implement 
\texttt{$\epsilon$GroundedFairSIRBot} in machine learning.
We believe that this algorithm can be implemented with neural networks, by substituting in learned strategies for the hard-coded parts of the \texttt{$\epsilon$GroundedFairSIRBot} algorithm that output
decisions (lines \ref{line:act_randomcoop}, \ref{line:conditional_coop}, \ref{line:trunc_type}).
Indeed, \citet{hutter2021learning} takes this approach to applying multi-agent reinforcement learning to programs in the class of \texttt{$\epsilon$GroundedFairBot}, of which \texttt{$\epsilon$GroundedFairSIRBot} is a generalization. 

\section{Discussion}
We have defined a new class of commitment
games that allow disclosure 
of private information conditioned on other players' 
commitments. 
Our folk theorem
shows that in these games,
efficient payoffs are always 
attainable in equilibrium,
which is not true in general without conditional disclosure
devices. 
Finally, we have provided an implementation of
this framework via robust program equilibrium,
which can be used by computer programs that read
each other's source code.

While conceptually simple, 
satisfying these
assumptions in practice requires a strong degree of mutual
transparency and conditional commitment ability, which is
not possessed by contemporary
human institutions or AI systems.
Thus, our framework represents an idealized standard for bargaining in the absence of a trusted third party, suggesting research priorities for the field of Cooperative AI
\cite{dafoe2020open}.
The motivation for work on this standard is that
AI
agents with increasing economic
capabilities, which would exemplify game-theoretic rationality
to a stronger degree than humans,
may be deployed in contexts where they make
strategic decisions on behalf of human principals \citep{geist}.
Given the potential for game-theoretically rational
behavior to cause cooperation failures \citep{myerson1983efficient,fearon1995rationalist},
it is important that such agents
are developed in ways that ensure 
they are able to cooperate effectively.

Commitment devices of this form would be particularly useful in 
cases where centralized institutions (\citet{dafoe2020open}, Section 4.4) for
enforcing or incentivizing cooperation
fail, or have not been constructed due to collective action problems.
This is because our devices do not require
a trusted third party,
aside from correlation signals.
A potential obstacle to the use
of these commitment devices is lack of
coordination in development of AI systems. This may lead to
incompatibilities in commitment device implementation, such 
that one agent cannot confidently verify that another's 
device meets its conditions for trustworthiness and hence type disclosure.
Given that commitments may be implicit in complex parametrizations of neural networks, 
it is not clear that independently trained agents will be 
able to understand each other's commitments without 
deliberate
coordination between developers. 
Our 
program equilibrium approach allows
for the relaxation of the coordination 
requirements needed to implement 
conditional information disclosure and commitment.
Coordination on target action profiles for commitment devices
or flexibility in selection of such profiles,
in interactions with multiple efficient and arguably ``fair'' profiles \citep{stastny2021normative},
will also be important for avoiding cooperation
failures due to equilibrium selection problems.




  %



\section*{Acknowledgments}

We thank Lewis Hammond
for helpful comments on this paper,
and thank Caspar Oesterheld both
for useful comments and for
identifying an important error in an earlier version of one of our proofs.



\bibliography{refs}

\begin{thebibliography}{31}
\providecommand{\natexlab}[1]{#1}

\bibitem[{Critch(2019)}]{critch2019parametric}
Critch, A. 2019.
\newblock A parametric, resource-bounded generalization of {L{\"o}b’s}
  theorem, and a robust cooperation criterion for open-source game theory.
\newblock \emph{The Journal of Symbolic Logic}, 84(4): 1368--1381.

\bibitem[{Dafoe et~al.(2020)Dafoe, Hughes, Bachrach, Collins, McKee, Leibo,
  Larson, and Graepel}]{dafoe2020open}
Dafoe, A.; Hughes, E.; Bachrach, Y.; Collins, T.; McKee, K.~R.; Leibo, J.~Z.;
  Larson, K.; and Graepel, T. 2020.
\newblock Open Problems in Cooperative {AI}.
\newblock arXiv:2012.08630.

\bibitem[{Dye(1985)}]{dye1985disclosure}
Dye, R.~A. 1985.
\newblock Disclosure of nonproprietary information.
\newblock \emph{Journal of accounting research}, 123--145.

\bibitem[{Fearon(1995)}]{fearon1995rationalist}
Fearon, J.~D. 1995.
\newblock Rationalist explanations for war.
\newblock \emph{International organization}, 49(3): 379--414.

\bibitem[{Forges(2013)}]{FORGES201364}
Forges, F. 2013.
\newblock A folk theorem for {Bayesian} games with commitment.
\newblock \emph{Games and Economic Behavior}, 78: 64--71.

\bibitem[{Geist and Lohn(2018)}]{geist}
Geist, E.; and Lohn, A.~J. 2018.
\newblock How might artificial intelligence affect the risk of nuclear war?
\newblock Rand Corporation.

\bibitem[{Grossman(1981)}]{grossman1981informational}
Grossman, S.~J. 1981.
\newblock The informational role of warranties and private disclosure about
  product quality.
\newblock \emph{The Journal of Law and Economics}, 24(3): 461--483.

\bibitem[{Grossman and Hart(1980)}]{grossman1980disclosure}
Grossman, S.~J.; and Hart, O.~D. 1980.
\newblock Disclosure laws and takeover bids.
\newblock \emph{The Journal of Finance}, 35(2): 323--334.

\bibitem[{Hagenbach, Koessler, and Perez-Richet(2014)}]{certifiable2014}
Hagenbach, J.; Koessler, F.; and Perez-Richet, E. 2014.
\newblock Certifiable Pre-play Communication: Full Disclosure.
\newblock \emph{Econometrica}, 82(3): 1093--1131.

\bibitem[{Hutter(2021)}]{hutter2021learning}
Hutter, A. 2021.
\newblock Learning in two-player games between transparent opponents.
\newblock arXiv:2012.02671.

\bibitem[{Jovanovic(1982)}]{jovanovic1982truthful}
Jovanovic, B. 1982.
\newblock Truthful disclosure of information.
\newblock \emph{The Bell Journal of Economics}, 36--44.

\bibitem[{Kalai et~al.(2010)Kalai, Kalai, Lehrer, and
  Samet}]{kalai2010commitment}
Kalai, A.~T.; Kalai, E.; Lehrer, E.; and Samet, D. 2010.
\newblock A commitment folk theorem.
\newblock \emph{Games and Economic Behavior}, 69(1): 127--137.

\bibitem[{Knott et~al.(2021)Knott, Venkataraman, Hannun, Sengupta, Ibrahim, and
  van~der Maaten}]{knott2021crypten}
Knott, B.; Venkataraman, S.; Hannun, A.; Sengupta, S.; Ibrahim, M.; and van~der
  Maaten, L. 2021.
\newblock {CrypTen}: Secure Multi-Party Computation Meets Machine Learning.
\newblock In Beygelzimer, A.; Dauphin, Y.; Liang, P.; and Vaughan, J.~W., eds.,
  \emph{Advances in Neural Information Processing Systems}.

\bibitem[{Kosba et~al.(2016)Kosba, Miller, Shi, Wen, and
  Papamanthou}]{smartcontract}
Kosba, A.; Miller, A.; Shi, E.; Wen, Z.; and Papamanthou, C. 2016.
\newblock Hawk: The Blockchain Model of Cryptography and Privacy-Preserving
  Smart Contracts.
\newblock In \emph{2016 IEEE Symposium on Security and Privacy (SP)}, 839--858.

\bibitem[{Kovenock, Morath, and Münster(2015)}]{Kovenock09informationsharing}
Kovenock, D.; Morath, F.; and Münster, J. 2015.
\newblock Information sharing in contests.
\newblock \emph{Journal of Economics \& Management Strategy}, 24: 570--596.

\bibitem[{LaVictoire et~al.(2014)LaVictoire, Fallenstein, Yudkowsky, Barasz,
  Christiano, and Herreshoff}]{lavictoire2014program}
LaVictoire, P.; Fallenstein, B.; Yudkowsky, E.; Barasz, M.; Christiano, P.; and
  Herreshoff, M. 2014.
\newblock Program Equilibrium in the Prisoner's Dilemma via {L{\"o}b's}
  Theorem.
\newblock In \emph{Workshops at the Twenty-Eighth AAAI Conference on Artificial
  Intelligence}.

\bibitem[{Lewis et~al.(2017)Lewis, Yarats, Dauphin, Parikh, and
  Batra}]{dealornodeal}
Lewis, M.; Yarats, D.; Dauphin, Y.~N.; Parikh, D.; and Batra, D. 2017.
\newblock Deal or No Deal? End-to-End Learning for Negotiation Dialogues.
\newblock arXiv:1706.05125.

\bibitem[{Martini(2018)}]{martini18}
Martini, G. 2018.
\newblock Multidimensional Disclosure.
\newblock
  \url{http://www.giorgiomartini.com/papers/multidimensional_disclosure.pdf}.

\bibitem[{Milgrom and Roberts(1986)}]{milgrom1986relying}
Milgrom, P.; and Roberts, J. 1986.
\newblock Relying on the information of interested parties.
\newblock \emph{The RAND Journal of Economics}, 18--32.

\bibitem[{Milgrom(1981)}]{milgrom1981good}
Milgrom, P.~R. 1981.
\newblock Good news and bad news: Representation theorems and applications.
\newblock \emph{The Bell Journal of Economics}, 380--391.

\bibitem[{Myerson and Satterthwaite(1983)}]{myerson1983efficient}
Myerson, R.~B.; and Satterthwaite, M.~A. 1983.
\newblock Efficient mechanisms for bilateral trading.
\newblock \emph{Journal of economic theory}, 29(2): 265--281.

\bibitem[{Oesterheld(2019)}]{oesterheld2019robust}
Oesterheld, C. 2019.
\newblock Robust program equilibrium.
\newblock \emph{Theory and Decision}, 86(1): 143--159.

\bibitem[{Oesterheld and Conitzer(2021)}]{oesterheld2021safe}
Oesterheld, C.; and Conitzer, V. 2021.
\newblock Safe {Pareto} Improvements for Delegated Game Playing.
\newblock In \emph{Proceedings of the 20th International Conference on
  Autonomous Agents and MultiAgent Systems}, 983--991.

\bibitem[{Okuno-Fujiwara, Postlewaite, and Suzumura(1990)}]{okuno1990}
Okuno-Fujiwara, M.; Postlewaite, A.; and Suzumura, K. 1990.
\newblock Strategic Information Revelation.
\newblock \emph{The Review of Economic Studies}, 57(1): 25--47.

\bibitem[{Peters and Szentes(2012)}]{Peters2012DefinableAC}
Peters, M.; and Szentes, B. 2012.
\newblock Definable and Contractible Contracts.
\newblock \emph{Econometrica}, 80: 363--411.

\bibitem[{Rubinstein(1998)}]{rubinstein}
Rubinstein, A. 1998.
\newblock \emph{Modeling Bounded Rationality}.
\newblock The MIT Press.

\bibitem[{Shin(1994)}]{shin1994burden}
Shin, H.~S. 1994.
\newblock The burden of proof in a game of persuasion.
\newblock \emph{Journal of Economic Theory}, 64(1): 253--264.

\bibitem[{Slantchev and Tarar(2011)}]{slantchev2011mutual}
Slantchev, B.~L.; and Tarar, A. 2011.
\newblock Mutual optimism as a rationalist explanation of war.
\newblock \emph{American Journal of Political Science}, 55(1): 135--148.

\bibitem[{Stastny et~al.(2021)Stastny, Riché, Lyzhov, Treutlein, Dafoe, and
  Clifton}]{stastny2021normative}
Stastny, J.; Riché, M.; Lyzhov, A.; Treutlein, J.; Dafoe, A.; and Clifton, J.
  2021.
\newblock Normative Disagreement as a Challenge for Cooperative {AI}.
\newblock arXiv:2111.13872.

\bibitem[{Tennenholtz(2004)}]{tennenholtz2004program}
Tennenholtz, M. 2004.
\newblock Program equilibrium.
\newblock \emph{Games and Economic Behavior}, 49(2): 363--373.

\bibitem[{Varian(2010)}]{varian2010computer}
Varian, H.~R. 2010.
\newblock Computer mediated transactions.
\newblock \emph{American Economic Review}, 100(2): 1--10.

\end{thebibliography}

}

\onecolumn

\appendix

\section{Other Conditions for Efficiency and Counterexamples}
\label{app:efficiency}

\subsection{Analysis of motivating example}
\label{app:sub:example}
Since country 2
can only either disclose both its strength $\theta$ and 
weak point
$\weakpoint$, or neither,
in our formalism of strategic information revelation
$\revelation(t_2) = \{\{t_2\}, T_2\}$.
If country 2 rejects the offer $1-s$,
players go to a war 
that country 2 wins with probability $p_W$ if
its army is weak, or
$p_S > p_W$ if strong.

Assume country 2 is strong and the prior probability of
a strong type is 
$q < \frac{p_S - p_W}{p_S - p_W + c_1 + c_2}$.
In the
perfect Bayesian equilibrium of $G$ (without type disclosure)
country 1 offers $1 - s = p_W - c_2$, which country 2 rejects \citep{slantchev2011mutual}.
That is, if country~1
believes country 2 is unlikely to be strong, country~1 makes 
the smallest
offer that
only a weak type would accept.
Thus with private information the countries
go to war and receive inefficient payoffs in equilibrium.
A strong country 2 also prefers not to disclose its type unconditionally.
Although this would guarantee
that country 1 best-responds with $1 - s = p_S - c_2$, which country 2 would accept,
given knowledge of the
weak point
$\weakpoint$ country 1 prefers to attack it and receive an extra~$\attackprofit$ payoff with certainty, costing $\exploited$ for country 2.
Country 2 would therefore be worse off by $\exploited$ than in equilibrium without
disclosure, where its expected payoff is $p_S - c_2$.

However, if country 2 
can
disclose its full type if
and only if country 1 commits to
$1-s = p_S$
that country~2 accepts,
and commits not to attack $\weakpoint$,
the countries can avoid war in equilibrium.
The profile $(1 - p(\theta), p(\theta))$ is not IC,
and hence cannot be achieved under
the assumptions of \citet{FORGES201364} alone,
because a weak country 2 would prefer
the strong type's payoff $p_S$
(absent type-conditional commitment by country 1).
In this example, conditional type disclosure is required
for efficiency due to a practical inability to
disclose military strength without also disclosing a
vulnerability 
(Table \ref{tab:sufficient}).
In other words, country 2's disclosure 
space $\revelation(t_2)$ is too restricted
for full unraveling to occur.
Interactions between advanced artificially
intelligent agents may feature similar problems
necessitating our framework.
For example,
if disclosure requires
sharing source code or the full set of
parameters of a neural network that lacks cleanly
separated modules, unconditional disclosure risks
exposing exploitable information. See also example 7 of
\citet{okuno1990}, in which full unraveling fails because a
firm does not want to disclose a secret technology
that provides a competitive advantage,
leading to inefficiency because other private information is not disclosed.

\subsection{Efficiency with unconditional disclosure}
\label{app:sub:other-sufficient}


\paragraph{Full unraveling} 
If full unraveling occurs in the base game~$G$, then the 
ability to conditionally disclose 
information becomes irrelevant. 
For example, consider
a modification of Example
1 
in
which there is no
weak point,
i.e., 
country $2$'s type is $t_2 = \theta$ rather than 
$t_2 = \{\theta, \weakpoint\}$.
A strong country~2 that can verify its strength to country 1
prefers to do so, since
this does not also help country 1 exploit it.
But because of this, if country 2 refuses to disclose its strength and
it is common knowledge that country 2 could verifiably disclose,
country 1 knows country 2 is weak.
Thus, all types are disclosed in equilibrium
without conditioning
on country 1's commitment.

Some sufficient and necessary conditions for full unraveling
have been derived.
\citet{certifiable2014} show that given
verifiable disclosure,
full unraveling is guaranteed if there are no cycles in
the directed graph defined by types that
prefer to pretend to be each other.
For full unraveling in
some classes of games
with multidimensional
types,
it is \textit{necessary} for one of the players' payoff to be
sufficiently nonconvex in the other's beliefs \cite{martini18}.
In Appendix \ref{app:unraveling}, we give an example where this condition fails,
thus
unconditional disclosure is insufficient even
without exploitable information.
However, even for games with full unraveling, the framework of
\citet{FORGES201364} is still insufficient for equilibria with non-IC payoffs, since that framework does not allow verifiable
disclosure (conditional or otherwise).

\paragraph{Partial disclosure and post-unraveling incentive compatibility} If 
in our original example country 2 could \textit{partially}
disclose its type, i.e., only the probability of winning a war but
not its
weak point,
conditional disclosure would not be necessary
(Table~\ref{tab:sufficient}).
This is because the strategy inducing the
efficient payoff profile $(1 - p(\theta) + c_2, p(\theta) - c_2)$ depends only on
the part of country 2's type that is disclosed by the unraveling argument.
Country 2 does not profit from lying about its exploitable, non-unraveled information — that is, the
payoff is IC with respect to that information, even if not IC in
the pre-unraveling game. Thus
country~1 does not need to know this information for the efficient payoff
to be achieved in equilibrium.
Formally, in this case $\revelation(t_2) = \partialrevelation(t_2)$, i.e., country 2 can choose
to disclose \textit{any} $\Theta_{2j} \ni t_2$,
producing an equilibrium of partial unraveling.
We can generalize this observation with the following
proposition.

\setcounter{proposition}{1}

\begin{proposition}
\label{partialrev}
Suppose that the devices in
$G(\mathcal{D})$ do not have
disclosure functions, and
$G$ is a game of strategic information revelation
with $\revelation(t_i) = \partialrevelation(t_i)$ for all $i, t_i$.
Let $q$ be updated to have support on
the subset $\Theta \subseteq T$ of types remaining
after unraveling.
As in \citet{FORGES201364}, assume
$\randomization$ is conditioned on $\fulltype$. Then a payoff profile $\mathbf{x}$ is achievable in a
Bayesian Nash equilibrium of $G(\mathcal{D})$ if and only if
it
is feasible, INTIR, and IC (with respect to the post-unraveling game and updated $q$).
\end{proposition}

\begin{proof}
This is an immediate corollary of Propositions 1 and~2 of \citet{FORGES201364},
applied to the base game induced by
unraveling (that is, with a prior updated on types being in the space $\Theta$).
\end{proof}

To our knowledge, it is an open question
which conditions are sufficient and necessary
for partial unraveling such that the efficient payoffs of
the post-unraveling game are IC.
An informal summary of Proposition \ref{partialrev} and characterizations of equilibrium payoffs
under our framework and that of \citet{FORGES201364} is: Given conditional
commitment ability,
efficiency can be achieved in equilibrium
if and only if there is sufficiently strong incentive compatibility,
conditional and verifiable disclosure
ability, or an intermediate combination of these
(see Table \ref{tab:sufficient}).

Proposition \ref{partialrev} is not vacuous; there exist games in which, given the
ability to partially, verifiably, and unconditionally disclose their private information,
players end up in an inefficient equilibrium
that is Pareto-dominated by a non-IC payoff.
Consequently, the alternatives to conditional 
information
disclosure
that we have considered are not sufficient to achieve
all feasible and INTIR payoffs even when partial disclosure is possible.
The game in Appendix \ref{app:unraveling} is
one example 
where such a payoff is efficient.
In the following example, the only efficient payoff
is IC. However, the set of equilibrium payoffs is smaller
than under our assumptions, and excludes some potentially 
desirable outcomes. For example, 
there is a non-IC $\epsilon$-efficient payoff that
improves upon the strictly efficient
payoff in utilitarian welfare (sum of all players' payoffs).

\begin{table*}[ht]
    \centering
    \begin{tabular}{|c|cc|}
    \hline
         & \textbf{Full} & \textbf{Partial} \\
         \hline
        \textbf{Conditional} & feasible, INTIR & feasible, INTIR \\
        \textbf{Unconditional} & feasible, INTIR, \{full unraveling or IC\} & feasible, INTIR, \{full unraveling or IC after unraveling\}\\
        \hline
    \end{tabular}
\caption{Sufficient conditions to achieve a payoff in a Bayesian Nash equilibrium of a commitment game, given different forms of verifiable type disclosure ability. ``Full'' means a player can only disclose their full type, while ``Partial'' means other subsets containing their type can be disclosed.}
\label{tab:sufficient}
\end{table*}

\setcounter{exmp}{1}
\begin{exmp}
[All-pay auction under incomplete information from 
\citet{Kovenock09informationsharing}]
Two firms $i=1,2$ participate in an
all-pay auction. Each firm has a private
valuation $s_i \stackrel{\text{iid}}{\sim} \text{Unif}[0,1]$ of a good. After
observing their respective valuations,
players simultaneously choose whether to disclose them.
Then they simultaneously submit bids $x_i \in [0, \infty)$,
and the higher bid wins the good,
with a tie broken by a fair coin.
Thus player $i$'s payoff is
$s_i(\mathbb{I}[x_i \geq x_{-i}] - \frac{1}{2}\mathbb{I}[x_i = x_{-i}]) - x_i$.
There is a
Bayesian Nash equilibrium of this base game in which
$x_i = s_i^2/2$,
and neither player discloses their valuation \citep{Kovenock09informationsharing}.
In this equilibrium, each player's ex interim payoff
is:
\begin{align*}
    \mathbb{E}_q u_i(\fulltype, (x_i, x_{-i})) &= s_iP(x_i \geq x_{-i}) - x_i \\
    &= s_iP(s_i^2/2 \geq s_{-i}^2/2) - s_i^2/2 \\
    &= s_i^2/2.
\end{align*}
The ex post payoffs are $(s_1 - s_1^2/2, -s_2^2/2)$ if $s_1 > s_2$, $(-s_1^2/2, s_2 - s_2^2/2)$ if $s_1 < s_2$, and $(s_1/2 - s_1^2/2, s_2/2 - s_1^2/2)$ otherwise.

Now, let $\varepsilon > 0$, and consider the following strategy $\corrpolicy$. For type profiles such that $s_1 > s_2$, let $x_1 = \min\{\varepsilon, s_1^3/2\}$ and $x_2 = 0$. For $s_2 > s_1$, let $x_1 = 0$ and $x_2 = \min\{\varepsilon, s_2^3/2\}$. Otherwise, let $x_1 = x_2 = 0$. Then:
\begin{align*}
    \mathbb{E}_q u_i(\fulltype, \corrpolicy) &= (s_i - \min\{\varepsilon, s_i^3/2\})P(s_i \geq s_{-i}) \\
    &= s_i^2 - s_i\min\{\varepsilon, s_i^3/2\} \\
    &\geq s_i^2/2.
\end{align*}
Thus the payoff induced by $\corrpolicy$ is feasible and INTIR,
because it exceeds the ex interim equilibrium payoff.
This is also an ex post Pareto improvement on the equilibrium,
because the ex post payoffs are $(s_1 - \min\{\varepsilon, s_1^3/2\}, 0)$
if $s_1 > s_2$, $(0, s_2 - \min\{\varepsilon, s_2^3/2\})$
if $s_2 > s_1$,
and $(s_1/2, s_2/2)$ otherwise.
Finally, this payoff is not IC, because
if $s_1 < s_2$, player 1 would profit
from $\corrpolicy$ conditioned on a type $s_1' \geq s_2$.

Note that the payoffs $(s_1, 0)$ and $(0, s_2)$, i.e., the case of $\varepsilon = 0$, are not
feasible. This non-IC payoff thus requires
$\varepsilon > 0$ and is inefficient by a margin
of $\varepsilon$.
However, in practice the players may favor this payoff
over $(s_1/2, s_2/2)$.
This is because the
non-IC payoff is
$\varepsilon$-welfare optimal, since whenever $s_i > s_{-i}$ for either
player, the supremum of the sum of payoffs is $s_i$.
\end{exmp}

\section{Partial Unraveling Example}
\label{app:unraveling}

Consider the following game of strategic information revelation. We will
show that in this game, there is a perfect Bayesian
equilibrium that is 
inefficient, and there is an 
efficient payoff profile that is not
IC. (That is, in this game, unconditional and partial type disclosure and the framework of
\citet{FORGES201364} are not sufficient to achieve efficiency.) This example is inspired by the
model in \citet{martini18}.

Player 2 is a village that lives around the base of a treacherous mountain (i.e., along the left and bottom sides of $[0,1]^2$). Their warriors are camped somewhere on the mountain, with coordinates $(\theta_x, \theta_y)$. Player 1 has no information on the warriors' location, hence the prior is $(\theta_x, \theta_y) \sim \text{Unif}[0,1]^2$.
But they know that warriors at higher altitudes are tougher; strength is proportional to $\min\{\theta_x, \theta_y\}$.
As in Example
1,
player 1 can offer a split $(s,1-s)$ of disputed territory.
If the players fight, then player 1 will send in paratroopers
at a location $(t_x, t_y)$
to fight player 2's warriors
at a cost proportional to their strength $\min\{\theta_x, \theta_y\}$.
They want to get as close as possible to minimize exposure to the elements, consumption of rations, etc (i.e., minimize the squared distance $(t_x - \theta_x)^2 + (t_y - \theta_y)^2$).
Meanwhile, player~2 wants the paratroopers to land as far from their village as possible, i.e., they want to maximize $\min\{t_x, t_y\}$.
Player 2 wins the ensuing battle with probability equal to their army's strength, i.e., $\min\{\theta_x, \theta_y\}$.

Formally, the game is as follows.
Only player 2 has private information, $(\theta_x, \theta_y) \sim \text{Unif}[0,1]^2$.
Player 2 has the unrestricted disclosure 
space $\revelation(t_2) = \partialrevelation(t_2)$. 
First, player~2 chooses $\Theta_{21} \in \revelation(t_2)$.
Then player 1 chooses $s \in [0,1]$. Player 2 can either accept or reject $s$. If player 2 accepts, the pair of payoffs is $(s, 1-s)$. Otherwise, player 1 plays $(t_x, t_y) \in [0,1]^2$, and for $c_1, c_2 > 0$:
\begin{align*}
    u_1 &= 1 - 2\min\{\theta_x, \theta_y\}
    - (t_x - \theta_x)^2 - (t_y - \theta_y)^2 - c_1 \\
    u_2 &= \min\{t_x, t_y\} + \min\{\theta_x, \theta_y\} - c_2.
\end{align*}
Let $f^+(x) = \max\{0, f(x)\}$ for any function $f$. Define:
\begin{align*}
    t^*(s) &= \begin{cases}
    \frac{3-\sqrt{5}}{2},& \text{if } s > c_2 - \frac{3-\sqrt{5}}{2} + 1\\
    \frac{(3-\sqrt{5})(c_2 - s - 1)}{2} + 1,& \text{otherwise}.
\end{cases}\\
    r(s) &= 1 - s - t^*(s) + c_2 \\
    q_{U_1(s)}(\theta_x, \theta_y) &= \frac{\mathbb{I}[\min\{\theta_x, \theta_y\} \leq t^*(s)]}{t^*(s)(2 - t^*(s))} \\
    q_{U_2(s)}(\theta_x, \theta_y) &= \frac{\mathbb{I}[\min\{\theta_x, \theta_y\} \in (r^+(s), t^*(s)]]}{(t^*(s) - r^+(s))(2 - t^*(s) - r^+(s))} \\
    p(s) &= \mathbb{I}[r(s) \geq t^*(s)] + \mathbb{I}[r(s) < t^*(s)] \cdot \frac{r^+(s)(2 - r^+(s))}{t^*(s)(2 - t^*(s))} \\
    m_1(s) &= \frac{(r^+(s)^4 - r^+(s)^3 - r^+(s)) - (t^*(s)^4 - t^*(s)^3 - t^*(s))}{3(t^*(s) - r^+(s))(2 - t^*(s) - r^+(s))} \\
    m_2(s) &= \frac{t^*(s)^2 - r^+(s)^2 - \frac{2}{3}(t^*(s)^3 - r^+(s)^3)}{(t^*(s) - r^+(s))(2 - t^*(s) - r^+(s))} \\
    s^* &= \argmax_s \{sp(s) + (1 - 2m_2(s) - 2(m_1(s) - t^*(s)^2) - c_1)(1-p(s))\} \\
    s(\theta_x, \theta_y) &= 1 - 2\min\{\theta_x, \theta_y\} + c_2.
\end{align*}

Then, we claim:

\begin{proposition}
Let $c_1 = c_2 = 0.1$. Let player 1's strategy be $s^*$ then $ (t^*(s), t^*(s))$ conditional on a given $s$ if player~2 does not disclose their type, otherwise $s(\theta_x, \theta_y)$ then $ (\theta_x, \theta_y)$. Let player 2's strategy be to disclose any and only types for which $\min\{\theta_x, \theta_y\} > t^*(s^*)$, and to accept any and only $s \leq 1 - t^*(s) - \min\{\theta_x, \theta_y\} + c_2$. Let player 1's belief update to $q_{U_1(s^*)}$ conditional on player 2 not disclosing their type, and to $q_{U_2(s)}$ conditional on player 2 not disclosing their type and rejecting $s$.

Then these strategies and beliefs are a perfect Bayesian equilibrium. Further, there exist $\theta_x, \theta_y$ such that this equilibrium is 
inefficient, and the payoff profile $(1 - \min\{\theta_x, \theta_y\} - t^*(s^*) + c_2, \min\{\theta_x, \theta_y\} + t^*(s^*) - c_2)$ is (1) a Pareto improvement on the equilibrium payoff and (2) not IC.
\end{proposition}

\begin{figure}[ht]
    \centering
    \includegraphics[width=8cm]{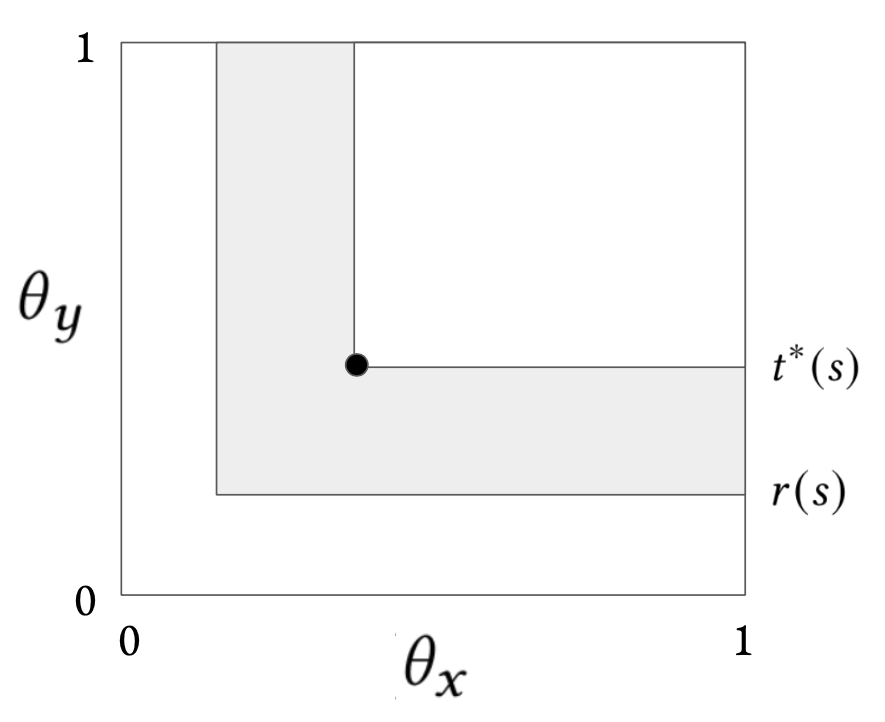}
    \caption{Player 2's type space, with the non-unraveled types conditional on a rejection of $s$ given in the shaded region. The black dot denotes $(\mathbb{E}_{q_{U_2(s)}} \theta_x, \mathbb{E}_{q_{U_2(s)}} \theta_y)$, player 1's optimal $(t_x, t_y)$ given a posterior $q_{U_2(s)}$ that is uniform on the shaded region. Unraveling stabilizes into this region because any types for which $\min\{\theta_x, \theta_y\} > \min\{\mathbb{E}_{q_{U_2(s)}} \theta_x, \mathbb{E}_{q_{U_2(s)}} \theta_y\}$ prefer to disclose, otherwise they prefer not to disclose, and types for which $\min\{\theta_x, \theta_y\} < r(s)$ prefer to accept $s$.
    }
    \label{fig:martini}
\end{figure}

\begin{proof}
We proceed by backward induction. If player 2 has not disclosed their type and has rejected $s$, then given beliefs $q_{U_2(s)}$, we solve for the optimal $(t_x, t_y)$. Player 1's expected payoff is $1 - 2\mathbb{E}_{q_{U_2(s)}}(\min\{\theta_x, \theta_y\}) - \mathbb{E}_{q_{U_2(s)}}((t_x - \theta_x)^2 + (t_y - \theta_y)^2) - c_1$. The squared loss is minimized at $(\mathbb{E}_{q_{U_2(s)}} \theta_x, \mathbb{E}_{q_{U_2(s)}} \theta_y)$. This is equivalent to the average of the centers of rectangles whose union composes the region $\min\{\theta_x, \theta_y\} \in [r^+(s), t^*(s)]$ (see Figure \ref{fig:martini}), weighted by the areas of these rectangles, which can be shown to be:
\begin{align*}
    t_x &= t_y \\
    &= \frac{(1-r^+(s))(t^*(s) + r^+(s)) + 1 - t^*(s)^2}{2(2 - t^*(s) - r^+(s))} \\
    &= t^*(s).
\end{align*}
Thus $(t^*(s), t^*(s))$ is a best response. If player 2 has disclosed their type and rejected $s$, then player 1's payoff is maximized at $(t_x, t_y) = (\theta_x, \theta_y)$.

Next, player 2's best response to any $s$ is to accept if and only if the acceptance payoff exceeds the rejection payoff given player~1's strategy, that is, $1 - s \geq t^*(s) + \min\{\theta_x, \theta_y\} - c_2$.

Then, given beliefs $q_{U_1(s)}$ for each $s$, player 1's optimal $s$ if player 2 does not disclose is:
\begin{align*}
    s^* &= \argmax_s \mathbb{E}_{q_{U_1(s)}}(u_1 | \text{accept } s)P(\text{accept } s) + \mathbb{E}_{q_{U_2(s)}}(u_1 | \text{reject } s)(1-P(\text{accept } s)).
\end{align*}

Given player 2's strategy, $P(\text{accept } s) = P(1 - s \geq t^*(s) + \min\{\theta_x, \theta_y\} - c_2) = P(\min\{\theta_x, \theta_y\} \leq r(s))$. Since $q_{U_1(s)}$ is uniform on $\min\{\theta_x, \theta_y\} \leq t^*(s)$, this probability is given by the ratio of the areas of the regions $\min\{\theta_x, \theta_y\} \leq r^+(s)$ and $\min\{\theta_x, \theta_y\} \leq t^*(s)$. Thus $P(\text{accept } s) = p(s)$. We have:
\begin{align*}
    \mathbb{E}_{q_{U_2(s)}}(u_1 | \text{reject } s) &= 1 - 2\mathbb{E}_{q_{U_2(s)}}(\min\{\theta_x, \theta_y\}) - \mathbb{E}_{q_{U_2(s)}}((t^*(s) - \theta_x)^2 + (t^*(s) - \theta_y)^2) - c_1 \\
    &= 1 - 2\mathbb{E}_{q_{U_2(s)}}(\min\{\theta_x, \theta_y\}) - \mathbb{V}_{q_{U_2(s)}}(\theta_x) - \mathbb{V}_{q_{U_2(s)}}(\theta_y) - c_1.
\end{align*}

It can be shown (Lemma \ref{calculus}) that $\mathbb{E}_{q_{U_2(s)}}(\min\{\theta_x, \theta_y\}) = m_2(s)$ and $\mathbb{V}_{q_{U_2(s)}}(\theta_x) = \mathbb{V}_{q_{U_2(s)}}(\theta_y) = m_1(s) - t^*(s)^2$.
Therefore $s^*$ is of the form given above.

If player 2 discloses, in the analysis above we now have:
\begin{align*}
    \mathbb{E}_{q_{U_2(s)}}(u_1 | \text{reject } s) &= 1 - 2\min\{\theta_x, \theta_y\} - c_1 \\
    P(\text{accept } s) &= \mathbb{I}[1 - s \geq 2 \min\{\theta_x, \theta_y\} - c_2] \\
    \mathbb{E}(u_1) &= \begin{cases}
    s,& \text{if } s \leq 1 - 2\min\{\theta_x, \theta_y\} + c_2\\
    1 - 2\min\{\theta_x, \theta_y\} - c_1,& \text{otherwise}.
\end{cases}
\end{align*}

Thus $s(\theta_x, \theta_y) = 1 - 2\min\{\theta_x, \theta_y\} + c_2$ is optimal, since $\mathbb{E}(u_1)$ increases with $s$ up to $1 - 2\min\{\theta_x, \theta_y\} + c_2$, after which it drops to $1 - 2\min\{\theta_x, \theta_y\} - c_1 < 1 - 2\min\{\theta_x, \theta_y\} + c_2$.

It can be shown 
that $s^* = 1$,
and so $r(s^*) < 0$ and $1-s^* < t^*(s^*) + \min\{\theta_x, \theta_y\} - c_2$ for any type. Given these responses, if player 2 does not disclose their type, their payoff is $t^*(s^*) + \min\{\theta_x, \theta_y\} - c_2$. If player 2 discloses their type, since we have shown that $s(\theta_x, \theta_y) = 1 - 2\min\{\theta_x, \theta_y\} + c_2$, player 2's payoff is  $2\min\{\theta_x, \theta_y\} - c_2$, and so player 2 prefers to disclose if and only if $\min\{\theta_x, \theta_y\} > t^*(s^*)$.

Finally, by the above strategy for player 2's type disclosure, if player 2 does not disclose, to be
consistent player~1 must update to the uniform distribution on the region defined by $\min\{\theta_x, \theta_y\} \leq t^*(s^*)$. Thus $q_{U_1(s^*)}$ is consistent. If player 2 also rejects~$s^*$, player 1 knows that $1-s^* < t^*(s^*) + \min\{\theta_x, \theta_y\} - c_2$, that is, $\min\{\theta_x, \theta_y\} > r(s^*)$. Thus the updated belief is uniform on $\min\{\theta_x, \theta_y\} \in (r(s^*), t^*(s^*)]$, so $q_{U_2(s^*)}$ is consistent. This proves that the proposed strategy profile and beliefs are a perfect Bayesian equilibrium.

Given $s^* = 1$, all player 2 types reject $s^*$ (offered if player~2 does not disclose) in equilibrium, since $t^*(1) > c_2$. The equilibrium payoffs for any types that do not disclose are:
\begin{align*}
    u_1 &= 1 - 2\min\{\theta_x, \theta_y\}
    - (t^*(1) - \theta_x)^2 - (t^*(1) - \theta_y)^2 - c_1 \\
    u_2 &= t^*(1) + \min\{\theta_x, \theta_y\} - c_2.
\end{align*}

Consider the payoff profile $(s_P(\theta_x,\theta_y), 1-s_P(\theta_x,\theta_y)) = (1 - t^*(1) - \min\{\theta_x, \theta_y\} + c_2, t^*(1) + \min\{\theta_x, \theta_y\} - c_2)$, induced by the strategy profile in which player 1 offers $s = s_P(\theta_x,\theta_y)$ and player 2 accepts any $s \leq s_P(\theta_x,\theta_y)$. This is feasible because player 2 only discloses if $\min\{\theta_x, \theta_y\} \leq t^*(1)$, and $c_2 < t^*(1) < \frac{1}{2}$, so $s_P(\theta_x,\theta_y) \in (0, 1)$. For this to be a Pareto improvement on the perfect Bayesian equilibrium, it is sufficient that $\min\{\theta_x, \theta_y\} > t^*(1) - c_1 - c_2$. This payoff profile is not IC, because player 2's payoff increases with $\min\{\theta_x, \theta_y\}$, so any player 2 for which $\min\{\theta_x, \theta_y\} < t^*(1)$ can profit from the strategy profile above conditioned on a type $(\theta_x', \theta_y')$ such that $\min\{\theta'_x, \theta'_y\} > \min\{\theta_x, \theta_y\}$.
\multilinecomment{
\begin{align*}
    s_P(\theta_x,\theta_y) - u_1 &= 1 - t^*(1) - \min\{\theta_x, \theta_y\} + c_2 - (1 - 2\min\{\theta_x, \theta_y\} - (t^*(1) - \theta_x)^2 - (t^*(1) - \theta_y)^2 - c_1) \\
    &= - t^*(1) + c_2 + \min\{\theta_x, \theta_y\} + (t^*(1) - \theta_x)^2 + (t^*(1) - \theta_y)^2 + c_1
\end{align*}
}

\end{proof}

\begin{lemma} Given $q_{U_2(s)}$ as defined above, $\mathbb{E}_{q_{U_2(s)}}(\min\{\theta_x, \theta_y\}) = m_2(s)$ and $\mathbb{V}_{q_{U_2(s)}}(\theta_x) = \mathbb{V}_{q_{U_2(s)}}(\theta_y) = m_1(s) - t^*(s)^2$.
\label{calculus}
\end{lemma}

\begin{proof}
We have:
\begin{align*}
    \mathbb{E}_{q_{U_2(s)}}(\min\{\theta_x, \theta_y\}) &= \frac{\int_0^1 \int_0^1 \min\{\theta_x, \theta_y\} \mathbb{I}[\min\{\theta_x, \theta_y\} \in [r^+(s),t^*(s)]] d\theta_x d\theta_y}{(t^*(s) - r^+(s))(2 - t^*(s) - r^+(s))} \\
    &= \frac{\int_0^1 \int_0^{\theta_y} \theta_x \mathbb{I}[\theta_x \in [r^+(s),t^*(s)]] d\theta_x d\theta_y + \int_0^1 \int_{\theta_y}^1 \theta_y \mathbb{I}[\theta_y \in [r^+(s),t^*(s)]] d\theta_x d\theta_y}{(t^*(s) - r^+(s))(2 - t^*(s) - r^+(s))} \\
    &= \frac{\int_0^1 \int_{\min\{\theta_y,t^*(s), r^+(s)\}}^{\min\{\theta_y, t^*(s)\}} \theta_x  d\theta_x d\theta_y + \int_{r^+(s)}^{t^*(s)} \int_{\theta_y}^{1} \theta_y d\theta_x d\theta_y}{(t^*(s) - r^+(s))(2 - t^*(s) - r^+(s))} \\
    &= \frac{\frac{1}{2}\int_0^1 (\min\{\theta_y, t^*(s)\}^2 - \min\{\theta_y,t^*(s),r^+(s)\}^2) d\theta_y + \int_{r^+(s)}^{t^*(s)} \theta_y (1 - \theta_y) d\theta_y}{(t^*(s) - r^+(s))(2 - t^*(s) - r^+(s))} \\
    &= \frac{\frac{1}{2}\int_0^{t^*(s)} \theta_y^2 d\theta_y + \frac{1}{2}\int_{t^*(s)}^1 t^*(s)^2 d\theta_y - \frac{1}{2}\int_0^{r^+(s)} \theta_y^2 d\theta_y - \frac{1}{2}\int_{r^+(s)}^1 r^+(s)^2 d\theta_y}{(t^*(s) - r^+(s))(2 - t^*(s) - r^+(s))} \\
    &+ \frac{\frac{1}{2}(t^*(s)^2 - r^+(s)^2) - \frac{1}{3}(t^*(s)^3 - r^+(s)^3)}{(t^*(s) - r^+(s))(2 - t^*(s) - r^+(s))} \\
    &= \frac{\frac{1}{6}t^*(s)^3 + \frac{1}{2}(1-t^*(s)) t^*(s)^2  - \frac{1}{6}r^+(s)^3 - \frac{1}{2}(1-r^+(s)) r^+(s)^2}{(t^*(s) - r^+(s))(2 - t^*(s) - r^+(s))} \\
    &+ \frac{\frac{1}{2}(t^*(s)^2 - r^+(s)^2) - \frac{1}{3}(t^*(s)^3 - r^+(s)^3)}{(t^*(s) - r^+(s))(2 - t^*(s) - r^+(s))}\\
    &= m_2(s).
\end{align*}
We showed above that $\mathbb{E}_{q_{U_2(s)}} \theta_x = \mathbb{E}_{q_{U_2(s)}} \theta_y = t^*(s)$. Further, $\mathbb{V}_{q_{U_2(s)}} \theta_x = \mathbb{E}_{q_{U_2(s)}} \theta_x^2 - (\mathbb{E}_{q_{U_2(s)}} \theta_x)^2$, so:
\begin{align*}
    \mathbb{E}_{q_{U_2(s)}} \theta_x^2 &= \frac{\int_0^1 \int_0^1  \theta_x^2 \mathbb{I}[\min\{\theta_x, \theta_y\} \in [r^+(s),t^*(s)]] d\theta_x d\theta_y}{(t^*(s) - r^+(s))(2 - t^*(s) - r^+(s))} \\
    &= \frac{\int_0^1 \int_0^1  \theta_x^2 \mathbb{I}[\min\{\theta_x, \theta_y\} \geq r^+(s)] d\theta_x d\theta_y - \int_0^1 \int_0^1  \theta_x^2 \mathbb{I}[\min\{\theta_x, \theta_y\} \geq t^*(s)] d\theta_x d\theta_y}{(t^*(s) - r^+(s))(2 - t^*(s) - r^+(s))} \\
    &= \frac{\int_{r^+(s)}^1 \int_{r^+(s)}^1  \theta_x^2 d\theta_x d\theta_y - \int_{t^*(s)}^1 \int_{t^*(s)}^1  \theta_x^2 d\theta_x d\theta_y}{(t^*(s) - r^+(s))(2 - t^*(s) - r^+(s))} \\
            &= \frac{\frac{1}{3}(1 - r^+(s))(1 - r^+(s)^3) - \frac{1}{3}(1 - t^*(s))(1 - t^*(s)^3)}{(t^*(s) - r^+(s))(2 - t^*(s) - r^+(s))} \\
            &= m_1(s).
\end{align*}
\end{proof}

\section{Proof of Proposition 1}
\label{app:prop_proof}

Let $\gencorr$ be the strategy profile of $(r^i_{d_i(t_i)}(\fulldevice_{-i},\rand))_{i=1}^n$. Then by hypothesis
$x_i(t_i) = \expectation{\fulltype_{-i}}{q}{t_i}{u_i(\fulltype, \gencorr)}$,
so $\mathbf{x}$ is feasible. Suppose that for some player $j$, for all correlated
policies~$\punishment_{-j}$
there exists a type $t_j$ such that:
\begin{align*}
    x_j(t_j) &<
    \max_{\act_j \in \mathcal{A}_j} \expectation{\fulltype_{-j}}{q}{t_j}{u_j(\fulltype,(\act_j,
\punishment_{-j}(\cdot | \fulltype_{-j})))}.
\end{align*}
Let
$\act^*_j = \argmax_{\act_j} \expectation{\fulltype_{-j}}{q}{t_j}{u_j(\fulltype,(\act_j,
\punishment_{-j}(\cdot | \fulltype_{-j})))}$.
Then if player $j$ with type $t_j$ deviates to $d'_j$ such that $r^j_{d'_j(t_j)}(\fulldevice_{-j},\rand) = \act^*_j$:
\begin{align*}
    &\mathbb{E}_{\fulltype_{-j} \sim q(\cdot | t_j)}(u_j(\fulltype, (\act^*_j,\gencorr_{-j}))|d'_j,\fulldevice_{-j}) \\
    &\qquad = \expectation{\mathbf{t}_{-j}}{q}{t_j}{u_j(\fulltype,(\act_j^*,\gencorr_{-j}(\cdot | \fulltype_{-j})))} \\
    &\qquad > x_j(t_j).
\end{align*}
This contradicts the assumption that $\mathbf{x}$ is the payoff of a Bayesian Nash equilibrium, therefore $\mathbf{x}$ is INTIR.

\section{Proof of Theorem 2}
\label{app:proof}

Fix the programs of players $i \neq j$ as $\fullprog_{-j}$.
\textbf{Suppose player $j$ uses $\prog_j$}.
Given this assumption, we omit the subscripts of $\progy$ and $\progr$.
Let $\progyrec{L}$ and $\progrrec{L}$ respectively
denote calls to $\progy$ and $\progr$
made at level $L$.
If $U_L < \epsilon$ and $U_{L+1} < \epsilon$ for some $L$ reached in
the call stack,
then every call to $\progyrec{L}$ and $\progyrec{L+1}$
immediately returns $\mathbf{1}$.
Consequently, every call to $\progrrec{L}$,
which must be a parent call to $\progyrec{L+1}$,
returns $\gencorri^{\rand}(\fulltype)$ because
line
15
in \texttt{$\epsilon$GroundedFairSIRBot}$^r$
evaluates to \texttt{True}.
(Notice that the shared random variables $\{U_L\}$ are essential --- if the programs
unconditionally cooperated using
independently sampled variables, an 
exponentially increasing number of
variables would each need to be less than
$\epsilon$ for all calls at a given level
to return the cooperative output.)
Let~$\mathcal{U}_L$ be the event that $U_L < \epsilon$ and $U_{L+1} < \epsilon$,
and $\mathcal{U}'_K$ be the event that $U_{2K-1} < \epsilon$ and $U_{2K} < \epsilon$.
Thus for the program profile to terminate in
finite time, it is sufficient
to show that with probability 1 there exists a finite $L$
such that $\mathcal{U}_L$ holds.
Given that $\mathcal{U}'_K$ for $K=1,2,\dots$ are independent,
because they do not overlap,
we have:

\begin{align*}
    P(\cap_{L=1}^\infty \mathcal{U}_L^c) &\leq P(\cap_{K=1}^\infty {\mathcal{U}'_K}^c) \\
    &= \prod_{K=1}^\infty (1-P(\mathcal{U}'_K)) \\
    &= \lim_{N \to \infty} (1-\epsilon^2)^{N} \\
    &= 0.
\end{align*}
Since $\cap_{L=1}^\infty \mathcal{U}_L^c$ is
the complement of the event we wanted to
guarantee, this proves termination
with probability 1.
Further, the event $\mathcal{U}_L$ is sufficient
for every call of $\progyrec{L-1}$ and $\progrrec{L-1}$
to return $\mathbf{1}$ and $\gencorri^{\rand}(\fulltype)$, respectively,
and this holds for all levels less than $L$.
Therefore all base calls of the programs in the proposed profile
return the corresponding $\gencorri^{\rand}(\fulltype)$ with probability 1.

Now \textbf{suppose player $j$ uses $\prog_j' \neq \prog_j$}.
Let $L^*$ be the smallest finite level such that $U_{L^*} < \epsilon$, $U_{L^*+1} < \epsilon$, and $U_{L^*+2} < \epsilon$
(which exists with probability 1 by a similar argument
to that above).
Then all $\progiyrec{L^*}$ and $\progiyrec{L^*+1}$ for $i \neq j$ return $\mathbf{1}$.
Further, every $\progirrec{L^*}$ for $i \neq j$
calls the truncated programs $[\progky]$ for $k \neq i$, guaranteed
to terminate by definition, thus
$\progirrec{L^*}$ terminates with either
$\gencorri^{\rand}(\fulltype)$ or $\genpunishi^{\rand}$.
But because $U_{L^*+2} < \epsilon$ also guarantees
that $\progirrec{L^*+1}$ terminates,
all calls to the programs of $i \neq j$ made by
player~$j$'s programs terminate.
Thus all base calls of programs in this profile with one deviation
terminate
with probability~1.

We now consider the possible cases.
Suppose $U_1 \geq \epsilon$ and $U_2 \geq \epsilon$.
First, note that any players $i \neq j$ using $\genpunishi^{\rand}$
know each other's types.
To see this, note that all calls to $\progiyrec{L^*}$ for $i \neq j$ return $\mathbf{1}$.
So any call to $\progiyrec{L^*-1}$ for $i \neq j$
will disclose to player $k \neq i$ if either $\act_k = \genpunishk^{\rand}$ or $U_{L^*} < \epsilon$.
The second condition is satisfied by assumption.
Inductively applying this argument for
$L \leq L^* - 1$, note that if
$U_{L+1} \geq \epsilon$,
we will only have $\act_k \neq \genpunishk^{\rand}$ if $U_L < \epsilon$ (satisfying line 
6
of \texttt{$\epsilon$GroundedFairSIRBot}$^r$), but then this is sufficient to have $\progiyrec{L}$ return $\mathbf{1}$ (line 
21).
If player~$j$ does not disclose their type,
all players $i \neq j$ return $\genpunishi^{\rand}$.
Otherwise,
\texttt{$\epsilon$GroundedFairSIRBot}$^r$
proceeds to line 
8
for all players $i \neq j$.
If player $j$ plays $\gencorrj^{\rand}(\fulltype)$, then
all other players also play $\gencorri^{\rand}(\fulltype)$,
giving the target payoff profile. Otherwise,
all players $i \neq j$ return $\genpunishi^{\rand}$.
We therefore have that with probability at least $(1-\epsilon)^2$, all players $i \neq j$ use 
$\genpunishi^{\rand}$ whenever
the outputs of $p_j'$ do not match those of $p_j$.
Hence:
\begin{align*}
    \mathbb{E}_{\fulltype_{-j} \sim q(\cdot | t_j)}(u_j(\fulltype, \corrpolicy)|\prog_j',\fullprog_{-j}) &\leq (1 - (1-\epsilon)^{2}) \overline{u} + (1-\epsilon)^{2}\mathbb{E}_{\fulltype_{-j} \sim q(\cdot | t_j)}(u_j(\fulltype, \corrpolicy)|\prog_j,\fullprog_{-j}) \\
    &\leq \delta + \mathbb{E}_{\fulltype_{-j} \sim q(\cdot | t_j)}(u_j(\fulltype, \corrpolicy)|\prog_j,\fullprog_{-j}).
\end{align*}

\end{document}
